\documentclass[a4paper,fleqn,longmktitle]{elsarticle}

\usepackage{amsmath,amsfonts}
\usepackage{algorithmic}
\usepackage{algorithm}
\usepackage{array}
\usepackage{textcomp}
\usepackage{stfloats}
\usepackage{url}
\usepackage{verbatim}
\usepackage{graphicx}
\usepackage{hyperref}
\usepackage{makecell}
\usepackage{pdflscape}
\usepackage{adjustbox}
\usepackage{enumitem}
\usepackage{multirow}
\usepackage{pifont}
\usepackage{tabularx}
\usepackage{rotating}   % For sideways tables
\usepackage{tablefootnote}
\usepackage{booktabs}
\usepackage[acronym,nopostdot]{glossaries}
\usepackage{subcaption}
\usepackage{lipsum}     % Dummy text (optional)
\usepackage{booktabs}   % Better table formatting (optional)
\usepackage{amssymb}% http://ctan.org/pkg/amssymb
\usepackage{pifont}% http://ctan.org/pkg/pifont
\usepackage{xcolor}
\newcommand{\xmark}{\ding{55}}%

\usepackage[numbers]{natbib}
\newacronym{cnn}{CNN}{Convolutional Neural Network}
\newacronym{pad}{PAD}{Presentation Attack Detection}
\newacronym{pa}{PA}{Presentation Attack}
\newacronym{kyc}{KYC}{Know Your Customer}
\newacronym{pai}{PAI}{Presentation Attack Instrument}
\newacronym{pais}{PAIs}{Presentation Attack Instruments}
\newacronym{gan}{GANs}{Generative Adversarial Networks}
\newacronym{dm}{DM}{Diffusion Models}
\newacronym{gdpr}{GDPR}{General Data Protection Regulation}
\newacronym{vit}{ViT}{Vision Transformers}
\newacronym{hdr}{HDR}{High Dynamic Range}
\newacronym{ppi}{PPI}{Pixels Per Inch}
\newacronym{dpi}{DPI}{Dots Per Inch}
\newacronym{gimp}{GIMP}{GNU Image Manipulator Program}
\newacronym{bpcer}{BPCER}{Bona-fide Presentation Classification Error Rate}
\newacronym{apcer}{APCER}{Attack Presentation Classification Error Rate}
\newacronym{acer}{ACER}{Average Classification Error Rate}
\newacronym{far}{FAR}{False Acceptance Rate}
\newacronym{frr}{FRR}{False Rejection Rate}
\newacronym{eer}{EER}{Equal Error Rate}
\newacronym{det}{DET}{Detection Error Trade-off}
\newacronym{bce}{BCE}{Binary Cross Entropy}
\newacronym{ai}{AI}{Artificial Intelligence}
\newacronym{pvc}{PVC}{Poli-Vinyl Chloride}
\newacronym{mlp}{MLP}{Multi-Layer Perceptron}
\newacronym{nlp}{NLP}{Natural Language Processing}
\newacronym{rnn}{RNN}{Recurrent Neural Neworks}
\newacronym{mhsa}{MHSA}{Multi-Head Self-Attention}
\newacronym{ocr}{OCR}{Optical Character Recognition}
\newacronym{id}{ID}{Identity Document}
\newacronym{ids}{IDs}{Identity Documents}

\begin{document}
\let\WriteBookmarks\relax
\def\floatpagepagefraction{1}
\def\textpagefraction{.001}

% Short title
%\shorttitle{}    

% Short author
%\shortauthors{}  

% Main title of the paper
\title{Privacy-Aware Detection of \\ Fake Identity Documents: Methodology, Benchmark, and Improved Algorithms (FakeIDet2)}  

% Title footnote mark
% eg: \tnotemark[1]
%\tnotemark[1] 

% Title footnote 1.
% eg: \tnotetext[1]{Title footnote text}
%\tnotetext[1]{} 

% First author
%
% Options: Use if required
% eg: \author[1,3]{Author Name}[type=editor,
%       style=chinese,
%       auid=000,
%       bioid=1,
%       prefix=Sir,
%       orcid=0000-0000-0000-0000,
%       facebook=<facebook id>,
%       twitter=<twitter id>,
%       linkedin=<linkedin id>,
%       gplus=<gplus id>]

\author{Javier Muñoz-Haro}%[orcid=0009-0004-5411-5492]
%\cormark[1] % Corresponding author indication
%\ead{javier.munnoz@uam.es} % Email id of the first author

\author{Ruben Tolosana}%[orcid=0000-0002-9393-3066]
%\cormark[2]
% Email id of the second author
%\ead{ruben.tolosana@uam.es}

\author{Julian Fierrez}%[orcid=0000-0002-6343-5656]
% Email id of the fifth author
%\ead{julian.fierrez@uam.es}

\author{\\Ruben Vera-Rodriguez}%[orcid=0000-0002-6338-8511]
% Email id of the third author
%\ead{ruben.vera@uam.es}

\author{Aythami Morales}%[orcid=0000-0002-7268-4785]
% Email id of the fourth author
%\ead{aythami.morales@uam.es}

% Corresponding author text
%\cortext[1]{Corresponding author}
%\cortext[2]{Principal corresponding author}

% Address/affiliation
%\affiliation{organization={Biometrics and Data Pattern Analytics Lab, Universidad Autónoma de Madrid},
            %addressline={Ciudad Universitaria de Cantoblanco}, 
            %citysep={Madrid}, % Uncomment if no comma needed between city and postcode
            %postcode={28049}, 
            %state={Madrid},
            %country={Spain}}

\address{Biometrics and Data Pattern Analytics Lab, Universidad Autónoma de Madrid, Ciudad Universitaria de Cantoblanco, Madrid, 28049, Madrid, Spain}

% For a title note without a number/mark
%\nonumnote{}

% Here goes the abstract
\begin{abstract}
    Remote user verification in Internet-based applications is becoming increasingly important nowadays. A popular scenario for it consists of submitting a picture of the user's Identity Document (ID) to a service platform, authenticating its veracity, and then granting access to the requested digital service. An ID is well-suited to verify the identity of an individual, since it is government issued, unique, and nontransferable. However, with recent advances in Artificial Intelligence (AI), attackers can surpass security measures in IDs and create very realistic physical and synthetic fake IDs. Researchers are now trying to develop methods to detect an ever-growing number of these AI-based fakes that are almost indistinguishable from authentic (bona fide) IDs. In this counterattack effort, researchers are faced with an important challenge: the difficulty in using real data to train fake ID detectors. This real data scarcity for research and development is originated by the sensitive nature of these documents, which are usually kept private by the ID owners (the users) and the ID holders (e.g., government, police, bank, etc.). The present study proposes a new privacy-aware methodology for research and development in fake ID detection that promotes collaboration between ID holders and AI researchers. In practice, the main contributions of our study are: 1) We present and discuss our proposed patch-based methodology to preserve privacy in fake ID detection research. 2) We provide a new public database, FakeIDet2-db, comprising over 900K real/fake ID patches extracted from 2,000 ID images, acquired using different smartphone sensors, illumination and height conditions, etc. In addition, three physical attacks are considered: print, screen, and composite. 3) We present a new privacy-aware fake ID detection method, FakeIDet2, which introduces two novel learnable modules: Patch Embedding Extractor and Patch Embedding Fusion. 4) We release a standard reproducible benchmark that considers physical and synthetic attacks from popular databases in the literature. The results achieved by our proposed FakeIDet2 in detecting very realistic fake IDs under unseen type of attacks are encouraging: 8.90\% and 13.84\% \gls{eer} in the very challenging datasets DLC-2021 and KID34K, respectively. FakeIDet2-db and the accompanying benchmark are publicly available\footnote{\url{https://github.com/BiDAlab/FakeIDet2-db}}.  
\end{abstract}  

% Use if graphical abstract is present
%\begin{graphicalabstract}
%\includegraphics{}
%\end{graphicalabstract}

% Research highlights
%\begin{highlights}
%\item 
%\item 
%\item 
%\end{highlights}

%\nocite{*}

% Keywords
% Each keyword is seperated by \sep
%\begin{keywords}
% \sep Benchmark \sep Database \sep Patch \sep Fake ID documents \sep Foundation Models
%\end{keywords}

\maketitle

% Main text  
\section{Introduction}
\label{sec:intro}

Analyzing the veracity of digital information is one of the great challenges facing society today \cite{Aslett2024, med_for_overview}. With the great advances made in the field of Generative AI, it is possible to synthesize non-existent content or to modify existing content \cite{Melzi_2023_ICCV,2020_JSTSP_GANprintR_Neves,rathgeb2022handbook}, using simple and fast tools easily available on the Internet. These methods can be used for good purposes, for example, correct biases \cite{pena25bias} or improve performance in some scenarios where data scarcity is present \cite{MELZI2024102322, sfdr_comp}, but they have their downsides, as they can also be used for malicious purposes, such as creating DeepFakes that are harmful \cite{2022_EAAI_DeepFakes_Tolosana, rathgeb2022handbook} or misinformation \cite{misinfo_detection}. Evidence of the latter has been found in recent news, where tampered documents\footnote{\href{https://www.nytimes.com/2025/02/13/nyregion/students-high-tech-fake-ids.html}{https://www.nytimes.com/2025/02/13/nyregion/students-high-tech-fake-ids.html}} or synthetic\footnote{\href{https://www.404media.co/inside-the-underground-site-where-ai-neural-networks-churns-out-fake-ids-onlyfake/}{https://www.404media.co/inside-the-underground-site-where-ai-neural-networks-churns-out-fake-ids-onlyfake/}} were used for the purchase of underage alcohol or to bypass remote verification systems such as \gls{kyc} to access digital services such as crypto exchanges or digital banking.

\begin{figure}[t]
    \centering
    \includegraphics[width=\linewidth]{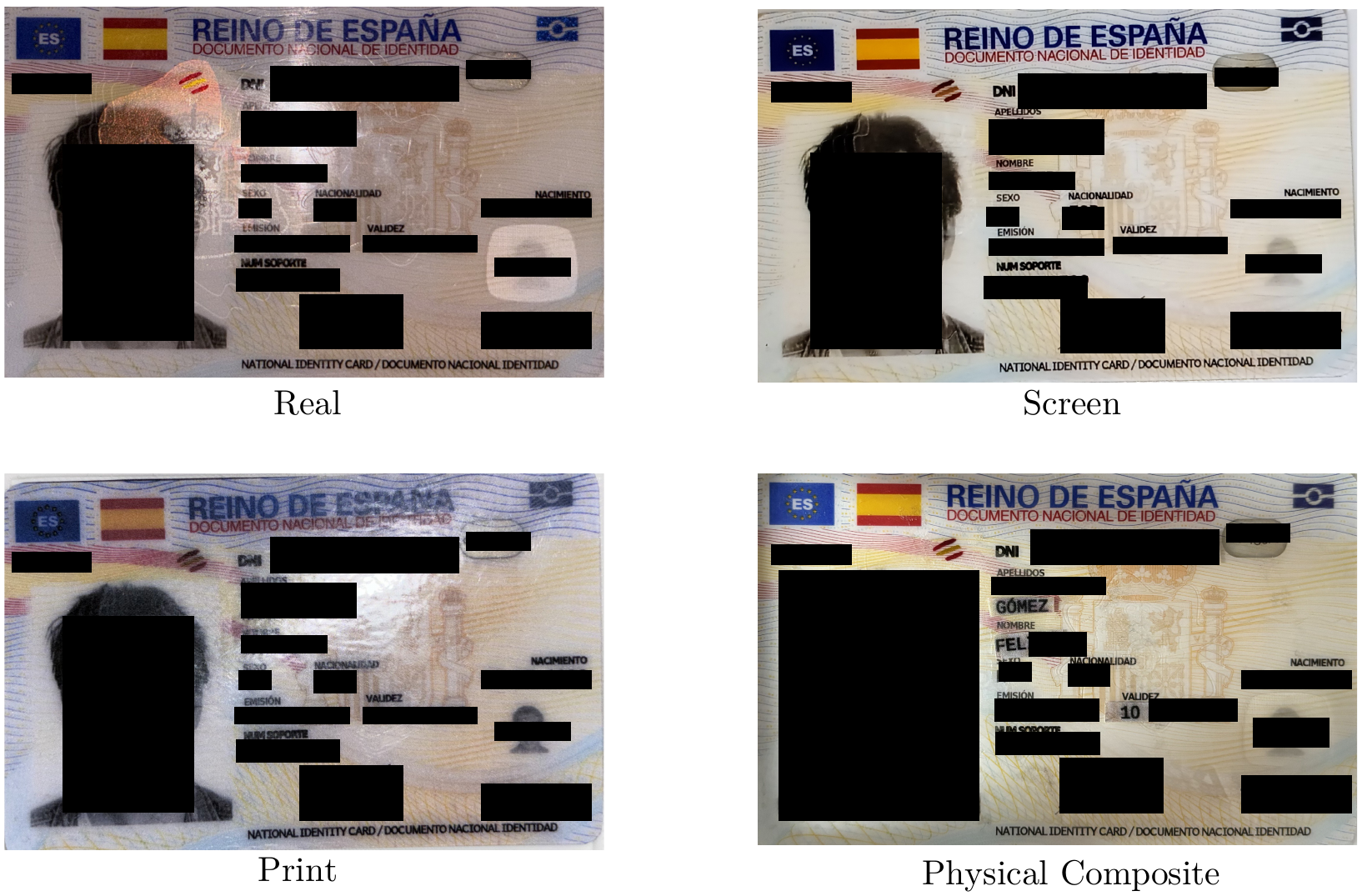}
    \caption{Examples of a real \gls{id} and popular physical attacks (print, screen, and physical composite) available in our proposed FakeIDet2-db. The personal information is redacted in this figure to protect the privacy of the subject.}
    \label{fig:id_examples}
\end{figure}

In order to advance in this challenging topic, the present study focuses on \gls{pad} \cite{2023_IJCB_SynFacePAD2023_Fang,2023_Book-PAD_Face_JHO} on \gls{ids}\footnote{Other related works define ID as ``Identity". Here we define ID as ``Identity Document".}. Concretely, as can be seen in Fig. \ref{fig:id_examples}, we focus on the most popular physical attacks in real-world scenarios: \textit{i) print} attacks are physical fake \gls{ids} created using an image or a scanned \gls{id}, which is printed on glossy paper to resemble the texture and appearance of real \gls{ids}; \textit{ii) screen} attacks, which consist in taking a picture of an \gls{id} displayed on a digital screen such as a laptop, tablet, or smartphone; and \textit{iii) composite} attacks, which do not require to tamper an entire document, but only small portions, such as date of birth, family name or the portrait image from the \gls{id} owner. In the particular case of \textit{physical composite} attacks, the impostor first prints fields with fake information on colored paper that resembles the colors of the target \gls{id}, and then crops each field. After that, these crops are lied on the \gls{id} document, occluding the original information with the fake one. All these kinds of attacks, a.k.a. \gls{pais}, are currently being used by attackers to surpass \gls{kyc} procedures, which explicitly require capturing live images of the user \gls{id} using a smartphone camera, granting them access to digital services using fake identities.

Several studies have preliminarily analyzed the problem of fake \gls{id} detection, proposing very valuable ideas and resources \cite{synth_id_card_db, bulatov2020midv, gonzalez2025forged}. However, there are several limitations in the field that need to be covered in a proper way to advance this research direction. First, there are no public databases with official (a.k.a. bona fide) \gls{ids} acquired under variable conditions, nor a standard benchmark that evaluates the performance of fake \gls{id} detection systems in real-world scenarios. So far, all public databases in the literature consider ``real" \gls{ids} some created by researchers under laboratory conditions, not official governments. From now on, we refer to these non-official \gls{ids} created by research groups as ``real" (with quotation marks). As a result, these ``real" \gls{ids} lack important information and patterns such as watermarks, not representing therefore a fidelity scenario of the problem. Some efforts have recently been made in this direction through the organization of international challenges such as DeepID\footnote{\href{https://deepid-iccv.github.io/}{https://deepid-iccv.github.io/}} (although the evaluation was private and is not publicly available yet) highlighting a considerable gap in performance between using ``real'' \gls{ids} and the official ones (0.99 vs 0.71 F1 score for the winner of the challenge). The main reason for considering this ``real" scenario is due to privacy concerns, as some fields of the \gls{ids} contain very sensitive information. These observations from state-of-the-art research trigger our \emph{key motivation for the present study: to research privacy-aware methods where more realistic data than the commonly used ``real" data can be used to develop better fake ID detectors based on machine learning, while keeping the data handling as private as possible.} 

\begin{figure}[t]
    \hspace*{-2.1cm}
    \centering
    \includegraphics[width=1.4\linewidth]{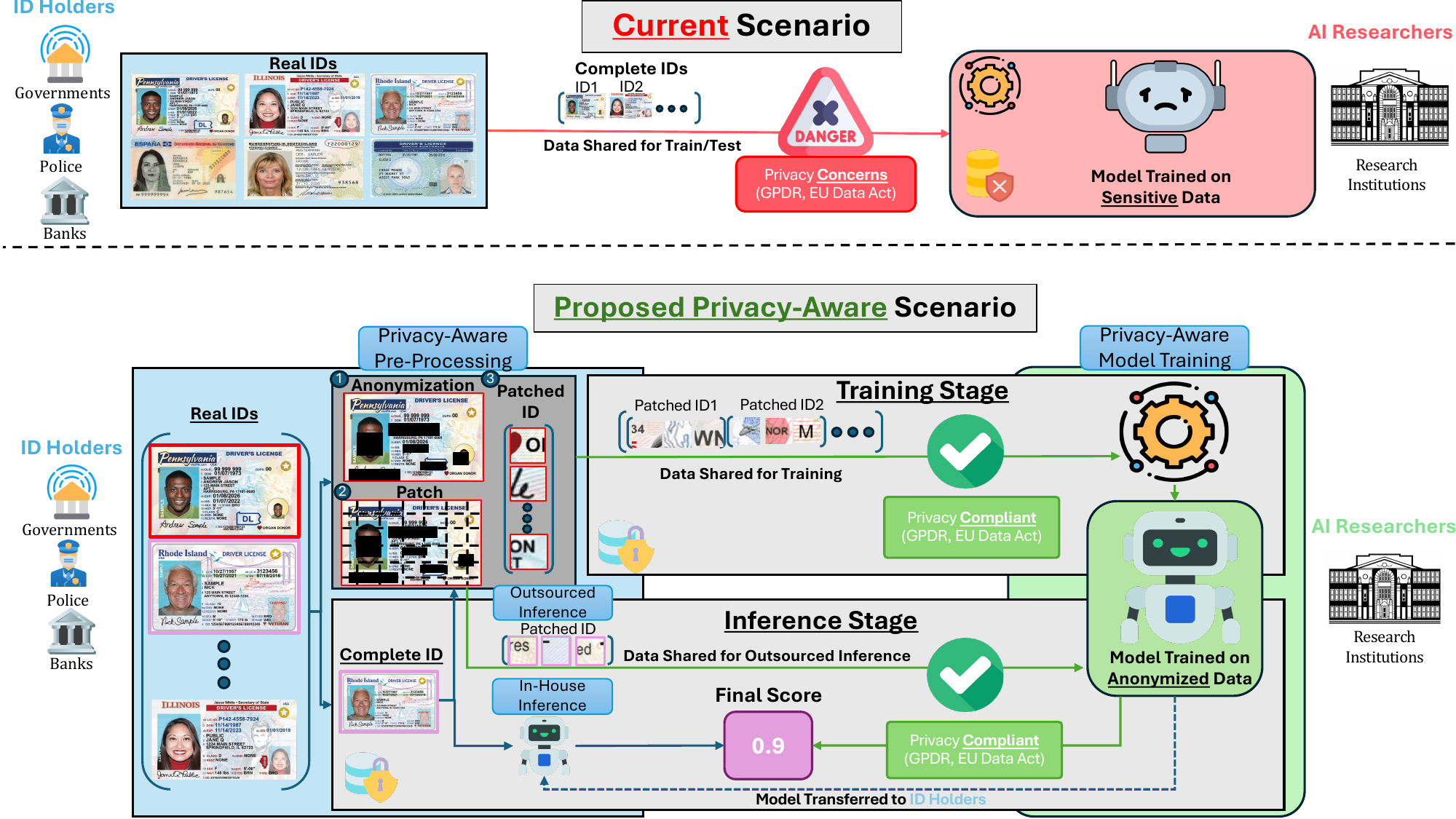}
    \caption{Our proposed framework for privacy-aware fake \gls{id} detection. The current scenario (top) raises privacy concerns regarding data regulations, and the need for complete \gls{id} sharing between \gls{id} Holders and AI Researchers. Our framework proposes a scenario where collections of non-ordered patches extracted from different levels of anonymized \gls{ids} can be used to preserve the sensitive information of \gls{id} owners during data sharing for training and inference in both in-house and outsourced schemes.}
    \label{fig:graph-abs}
\end{figure}

In the present study, we hypothesize that using \textit{patches} from \gls{ids}, instead of the whole \gls{id}, to train fake \gls{id} detectors might provide benefits from both the performance and privacy points of view in real-world scenarios. Fig.~\ref{fig:graph-abs} provides a graphical representation of the current scenario considered today for the problem of fake \gls{id} detection (top), versus our proposed privacy-aware scenario (bottom). In both scenarios, there are two main actors: \textit{\gls{id} Holders} (e.g., government, police, bank, etc.), which provide digital services to citizens and own large-scale datasets of real \gls{ids}, and the \textit{AI Researchers} (e.g., research institutions), which have the experience and technology to develop fake \gls{id} detectors, but not the data. The current scenario based on sharing the whole \gls{id} raises privacy concerns, limiting the development of robust fake \gls{id} detectors. Also, there are important concerns in terms of attacks, e.g., if the attackers are able to access the model trained on whole \gls{id} private data, they could use techniques that retrieve training data with small prior information about the data distribution \cite{alcala25iccv,pmlr-v235-feng24h, Zhang_2020_CVPR}. 

Our proposed privacy-aware scenario is based on the premise that an individual patch contains much less sensitive information than the whole ID. In addition, to provide flexibility to the particular restrictions of each \gls{id} Holder, we explore different anonymization levels, removing patches from sensitive regions as desired. For example, \gls{id} Holders may request AI Researchers to train a fake \gls{id} detector but they may not want to share any personal information about the \gls{id} owners. As a result, \gls{id} Holders can anonymize the sensitive sections from an \gls{id}, depending on their preferences for the task at hand, and extract patches from the areas that hold non-anonymized information. Additionally, to increase privacy, the extracted patches can be randomized regarding the order in which they are shared with the AI Researchers, further preventing reconstruction. This flexibility allows for fruitful collaboration between \gls{id} Holders and AI Researchers to train fake \gls{id} detectors in any \gls{id} Holder restrictions.

Once the fake \gls{id} detector is trained by the AI Researchers, the \gls{id} Holders can decide whether inference is carried out inside the AI Researchers infrastructures or internally, as seen in Fig.~\ref{fig:graph-abs}. For example, if \gls{id} Holders want to perform in-house inference using sensitive information that they only have access to, AI Researchers can transfer the fake \gls{id} detector previously trained using anonymized \gls{ids} to \gls{id} Holders. However, if \gls{id} Holders do not have such resources; they can rely on the AI Researchers for inference, since the \gls{id} Holders always control the amount of sensitive information shared for each \gls{id}. The proposed scenario can also be applicable to the research community in order to advance in this challenging topic, since research groups can share patches of \gls{ids}, including different levels of anonymization as desired.

The main contributions of the present study are as follows:

\begin{itemize}
    \item We propose and explore a new methodology to facilitate the collaboration between \gls{id} Holders and AI Researchers for advancing real-world fake \gls{id} detection. Concretely, instead of feeding the fake \gls{id} detectors with the whole \gls{id}, as is commonly done in the literature, we explore privacy-aware scenarios based on different patch sizes (64$\times$64 and 128$\times$128) and anonymization configurations (non-, pseudo-, and fully-anonymized \gls{id}). This proposal increases flexibility to both \gls{id} Holders and AI Researchers for both the training and inference stages, as shown in Fig.~\ref{fig:graph-abs}. Depending on the particular restrictions of the \gls{id} Holder, different setups can be selected to trade-off and optimize both privacy and performance. 

    \item We provide a new database, FakeIDet2-db, which contains official real data, along with \textit{print}, \textit{screen}, and \textit{physical composite} attacks. To our knowledge, FakeIDet2-db is the first database containing physical composite forgeries, as previous ones considered digital composites. In total, FakeIDet2-db comprises 922,057 patches extracted from 2,000 \gls{id} images (pseudo- and fully-anonymized configurations) from 47 different Spanish \gls{ids}. In addition, to increase the variability during the acquisition, we consider three smartphones for different end-users, including different hardware and software: iPhone 15 (high-end), Xiaomi Mi 9T Pro (medium-end) and Redmi 9C NFC (low-end). To add further variability, images are taken at 3 different heights (10 cms, 12.5 cms, and 15 cms) with respect to the \gls{id} and also different illumination conditions: \textit{no-light} with the camera flash activated, \textit{dim-light} with/without the camera flash activated, and \textit{bright-light} with/without the camera flash activated. 
    
    \item We propose a novel privacy-aware fake ID detector, FakeIDet2, which introduces two new learnable modules: the Patch Embedding Extractor and the Patch Embedding Fusion. The Patch Embedding Extractor leverages the DINOv2 foundation model to learn representations of individual patches from real/fake \gls{ids}, using an optimized AdaFace loss function \cite{kim2022adaface} through adaptable class weights, which evolve as training progresses \cite{fierrez18fusion}. Once the embeddings have been obtained, they are merged jointly in the Patch Embedding Fusion module. 
 
    \item We have developed a standard reproducible benchmark on which we evaluate the performance of our proposed FakeIDet2 in 3 different \gls{id} databases (DLC-2021 \cite{polevoy2022document}, KID34K \cite{kid34k}, and Benalcazar \textit{et al.} synthetic database of \cite{synth_id_card_db}), considering both physical and synthetic attacks, with the goal of assessing the performance in realistic out-of-distribution data. The proposed FakeIDet2-db and benchmark will be publicly available for reproducibility reasons in our GitHub repository\footnote{\url{https://github.com/BiDAlab/FakeIDet2-db}}, allowing a straightforward comparison of recent approaches with the state of the art. 
\end{itemize}

A preliminary analysis of this privacy-aware scenario was conducted in \cite{munoz2025exploring}. The present article significantly improves \cite{munoz2025exploring} in the following aspects: \textit{i)} We present our new publicly available FakeIDet2-db, which increases the number of unique real \gls{ids} (47 subjects) along with triple number of acquisition devices (now 3), illumination settings (now 3) and heights (now 3), resulting in 922,057 patches extracted from 2,000 \gls{id} images (pseudo- and fully-anonymized configurations). \textit{ii)} We include physical composite attacks in our novel FakeIDet2-db. To our knowledge, this is the first database that includes this kind of \gls{pai}. \textit{iii)} We propose a novel approach for fake \gls{id} detection, FakeIDet2, which overcomes the limitations in patch-level fusion encountered in FakeIDet~\cite{munoz2025exploring}. While FakeIDet performs a simple mean of scores assigned to each individual patch, FakeIDet2 comprises two learnable modules (i.e., Patch Embedding Extractor and Patch Embedding Fusion) to leverage the fusion of learned patch representations, depending on the visual characteristics from real/fake \gls{ids}. The representations at patch level are obtained through an optimized version of AdaFace \cite{kim2022adaface}, including dynamic class weights, which accounts for the severe class imbalance present in the databases for the different \gls{pais}. The fusion of the representations is performed through \gls{mhsa}, which ponders the embedding values based on the relations between them, to then compress them into a single embedding, through attention pooling based on \cite{attn_pool}. The impact of these improvements can be seen in the results (Sec. \ref{sec:exp_results}). And
\textit{iv)} In order to advance in this research field, we propose a standard public benchmark considering a wide variety of physical and synthetic attacks from different databases in the literature, allowing a direct comparison of recent approaches with the state of the art.

The remainder of the paper is organized as follows. Sec. \ref{sec:rel_work} provides an overview of the fake \gls{id} detection problem. In Sec. \ref{sec:proposed_db} we present our FakeIDet2-db, delving into the acquisition process for capturing real \gls{ids} and their fake counterparts. Sec. \ref{sec:proposed_method} introduces our novel privacy-aware fake ID detection method, FakeIDet2. Sec. \ref{sec:experiments} proposes a reproducible experimental protocol and standard benchmark, considering experiments in both intra- and cross-database scenarios. Finally, Sec. \ref{sec:exp_results} analyzes the performance of our proposed fake detector using the proposed experimental protocol and standard benchmark and Sec. \ref{sec:conclusion} draws the final conclusions.
\section{Related Work}
\label{sec:rel_work}

This section provides an overview of the fake \gls{id} detection problem. Sec. \ref{sub:databases} provides a revision of the databases considered in the field. Sec. \ref{sub:fakeDetectors} focuses on representative fake \gls{id} detectors and recent international challenges. Finally, we summarize key aspects in Sec. \ref{sub:summary}.  

\begin{table}
\resizebox{1.1\textwidth}{!}{
\hspace*{-1.2cm}
\centering
    \begin{tabular}{ccccccccc}
        \toprule
        \textbf{Database} & \textbf{\#Samples} & \textbf{\#IDs} & \textbf{\makecell{\#Devices}} & \textbf{\#PAIs} & \textbf{\makecell{Official \\ Real IDs?}} & \textbf{Public?} \\
        \midrule
        \makecell[c]{BID (2020) \\ \cite{soares2020bid}} & 28,800 & 28,800 & N/A & 8 & \textcolor{red}{\xmark} & \textcolor{green}{\checkmark} & \\
        \makecell[c]{DLC-2021 (2021) \\ \cite{polevoy2022document}} & 1,424 (videos) & 1,000 & 2 & 4 & \textcolor{red}{\xmark} & \textcolor{green}{\checkmark} \\
        \makecell[c]{Mudgalgundurao \textit{et al.} (2022) \\ \cite{mudgalgundurao2022pixel}} & 97,477 & 433 & N/A & 2 & \textcolor{green}{\checkmark} & \textcolor{red}{\xmark} & \\
        \makecell[c]{Benalcazar \textit{et al.} (2023) \\ \cite{synth_id_card_db}} & 3,000 & 3,000 & Synthetic & 1 & \textcolor{red}{\xmark} & \textcolor{green}{\checkmark} &  \\
        \makecell[c]{KID34K (2023) \\ \cite{kid34k}} & 34,662 & 92 & 12 & 3 & \textcolor{red}{\xmark} & \textcolor{green}{\checkmark} \\
        \makecell[c]{IDNet (2024) \\ \cite{idnet}} & 837,060 & 837,060 & Synthetic & 7 & \textcolor{red}{\xmark} & \textcolor{red}{\xmark} \\
        \makecell[c]{Open-Set (2024) \\ \cite{open-set}} & 5,424 & N/A & Synthetic & 2 & \textcolor{red}{\xmark} & \textcolor{green}{\checkmark} \\
        \makecell[c]{Gonzalez \textit{et al.} (2025) \\ \cite{gonzalez2025forged}} & 190,000 & 190,000 & N/A & 5 & \textcolor{green}{\checkmark} & \textcolor{red}{\xmark} \\
        \makecell[c]{FakeIDet-db (2025) \\ \cite{munoz2025exploring}} & 90 & 30 & 1 & 2 & \textcolor{green}{\checkmark} & \textcolor{green}{\checkmark}  \\
        \hline
        \makecell[c]{\textbf{FakeIDet2-db} \\ \textbf{(proposed)}} & 2,000 & 47 & 3 & 3 & \textcolor{green}{\checkmark} & \textcolor{green}{\checkmark} \\
        % Add more rows as needed
        \bottomrule
    \end{tabular}
    }
    \caption{Summary of databases considered in the literature for research in fake ID detection. We indicate whether the databases are available or not for public research.}
    \label{tab:db_summary}
\end{table}

\subsection{Databases}\label{sub:databases}

Table~\ref{tab:db_summary} provides a summary of the most relevant public/private databases in the field of fake \gls{id} detection. One of the first family of databases was the MIDV family \cite{arlazarov2019midv, Bulatov2022, bulatov2020midv}. With the original purpose of \gls{ocr}, Bulatov \textit{et al.} synthetically created a set of physical fake \gls{ids}, passports, and driving licenses. They used several digital templates from multiple countries that they filled with names and addresses from Wikipedia and artificially generated faces. As different versions of the database were released, the number of fake samples increased. The DLC-2021 database \cite{polevoy2022document} used the documents of the MIDV-family to create \gls{pai}s: \textit{color print}, \textit{gray print}, and \textit{screen}. The authors also provided ``real" samples, although they should be considered fake samples as they are just high-quality prints built by them, not official \gls{ids}. A similar work is presented in the KID34K dataset \cite{kid34k}, where Park \textit{et al.} introduced a dataset of 34,000 images from 82 Korean \gls{ids} and driving licenses. As in the MIDV family, the authors claimed to release ``real'' samples, but they were also built by the authors, not official documents. In order to prove the quality of the ``real'' samples, they trained a \gls{cnn} on the KID34K dataset using the whole \gls{id}. The evaluation was performed using official Korean \gls{ids} and ``real'' documents. After that, the authors applied a dimensionality reduction technique to visualize the model's embeddings in 2D, where they showed that embeddings from both official and ``real'' generated \gls{ids} from the KID34K database were similarly distributed in the embedding space. These results raise questions about the fake \gls{id} detection problem, since fake \gls{ids} can be generated with such a resemblance to real ones that deep-learning models may not be able to extract distinguishable features for proper \gls{pad}. Soares \textit{et al.} \cite{soares2020bid} introduced the Brazilian Identity Document (BID) database with a total of 28,800 images from Brazilian \gls{ids} that were digitally altered using different algorithms that removed blur and replace different personal information from the real \gls{ids} to comply with data privacy regulations. In \cite{gonzalez2025forged}, Gonzalez and Tapia proposed a more sophisticated attack procedure, i.e., printing the digital copy on a \gls{pvc} card, which resembles even more the appearance of real \gls{ids}.

Regarding synthetic data, Benalcazar \textit{et al.} proposed in \cite{synth_id_card_db} to use \gls{gan} to produce synthetic Chilean \gls{ids}. The proposed GAN was trained using only real Chilean \gls{id} samples. Although the data synthesized by the authors are not strictly real (and can be detected using GAN traces \cite{2020_JSTSP_GANprintR_Neves}), this was presented as a good idea to partially cover the lack of real data, for example, as a data enhancement strategy in training. Another synthetic database is IDNet \cite{idnet}, where Xie \textit{et al.} proposed a database that comprises synthetic copies of \gls{ids} from 10 countries and 10 states of the United States of America. Given an \gls{id} template, the portrait images and the owners' data were synthetically generated, with special care that the fields corresponded to the gender and age of the synthetic portrait image. In addition, the authors created six different types of digital manipulations, including inpainting \cite{2023_IRL-Net_Ahmad}, field cropping, and face morphing. A total of 837,060 \gls{ids} were generated, making it the largest and most diverse to date, although real data are not included. Markham \textit{et al.} \cite{open-set} developed Open-Set, a database that leverages neural style transfer, via \gls{gan}-based models to transfer textures from \textit{print} and \textit{screen} attacks to ``real" images from the MIDV-2020 and DLC-2021 databases, creating even more samples of attacks.

\subsection{Fake \gls{id} Detection Methods and International Challenges}\label{sub:fakeDetectors}

One of the first methods for the detection of fake IDs was presented in \cite{mudgalgundurao2022pixel}, where the authors trained a \gls{cnn} by pixel classification using an internal database, reporting a 2.22\% \gls{eer}. Gonzalez and Tapia proposed in \cite{gonzalez2025forged} a two-stage system which first used a neural network to evaluate if the \gls{id} is real or fake, evaluating digital PAIs of type \textit{composite} and \textit{synthetic}. The second network classified the \gls{id} between real or fake, evaluating physical PAIs such as \textit{print, display}, and \textit{PVC}. When concatenating both systems, the Bona-fide Presentation Classification Error Rate at a fixed decision threshold $\tau=0.01$ ($\textrm{BPCER}_{100}$) was 0.92\%.

A corroboration that the fake \gls{id} detection field is receiving increasing attention in recent years is the organization of international challenges at top conferences. The first challenge was introduced at IEEE IJCB 2024 \cite{tapia2024first}. The organizers did not provide data to prepare the competition, and the fake \gls{id} detectors submitted were evaluated on real \gls{ids}. The winner of the challenge achieved an \gls{eer} of 21.87\%. The second edition of this challenge has been hosted at IEEE IJCB 2025\footnote{\url{https://sites.google.com/view/ijcb-pad-id-card-2025/}}, where the organizers proposed two tracks. The first track considered \gls{ids} created by the organizers using \gls{pvc} cards as ``real" documents, the rest as attacks, and the second track treated only official \gls{ids} as real. Only for the first track, a database was given to the participants. The winner of the first track reported an \gls{eer} of 11.34\%, while for the second track, the best team reported an \gls{eer} of 6.36\%, but the second and third teams reported \gls{eer}s of 23.87\% and 31.94\%, respectively. These results show that this is a very challenging task, especially because of the lack of public real data. More recently, the first challenge including digital manipulations for fake \gls{id} detection (DeepID challenge) has been organized at IEEE/CVF ICCV 2025\footnote{\url{https://deepid-iccv.github.io/}}. This challenge proposed two tracks: one for binary classification focused on \gls{pad} (i.e., detecting fake and real documents) and one for localization (i.e., providing a mask showing the tampered sections). The organizers provided a database for all participants, named FantasyID, which contained fake templates from different countries from different regions of the world. The ``real'' \gls{id} data was created by printing high quality fakes using \gls{pvc} cards and taking images of them. After that, those images were digitally altered through face-swapping and text-inpainting methods. For the detection track, the evaluation was performed using the F1 score with a decision threshold fixed at $\tau = 0.5$ on two datasets: the FantasyID evaluation set, which contained \gls{pai}s that were not available in the training set, and a private dataset provided by the company PXL Vision, with the same type of attacks as in the FantasyID dataset, but using official \gls{ids} as real. The final score, named Aggregated F1 score, was a pondered mean that gave more importance to the F1 score obtained over the private dataset, hence benefiting models that generalized better to out-of-distribution data. The winning team achieved an Aggregated F1 score of 0.8, obtaining an F1 score of 0.99 and 0.72 in both the FantasyID dataset and the private dataset, respectively. These results, although impressive, show that there is a present issue in domain adaptation between official real \gls{ids} and fake samples created under laboratory conditions that are considered ``real''. Hence, providing privacy-aware mechanisms that allow sharing more realistic ID data without compromising individual's personal information is fundamental to further advance in this research field.

\subsection{State of the Art: Summary}\label{sub:summary}

 A clear trend can be observed on the basis of previous sections. Public databases in the literature do not have official real data available, and fake \gls{id} detectors trained using private databases with real data do not release the code or the weights to the research community, making it impossible to perform fair comparisons among state-of-the-art methods. Moreover, the characteristics of these databases are not always detailed (e.g., acquisition devices or number of unique IDs), which makes it even more difficult to advance in the field. The present study aims to advance through the following key contributions: \textit{i)} the release of a new database, FakeIDet2-db, increasing the number of images from real and fake \gls{ids}, \textit{ii)} the inclusion of a new kind of \gls{pais}, named \textit{physical composite}, to the proposed database, being the first one that contains this kind of attack, \textit{iii)} a novel method for privacy-aware fake \gls{id} detection based on fusion of representations from individual patches, and \textit{iv)} a new standard public benchmark that considers a wide variety of physical and synthetic attacks included in other public databases in the literature, considering \gls{ids} with different templates and demographic regions.
\section{Proposed Database: FakeIDet2-db}
\label{sec:proposed_db}

This section describes our new FakeIDet2-db. Concretely, Sec.~\ref{sub:bona_fide_id} and Sec.~\ref{sub:attacks_id} provide all the details of the real and fake \gls{id} acquisition, whereas Sec.~\ref{sub:fromid2patch} describes the post-processing carried out for the different anonymization and patch size configurations.   

\subsection{Acquisition Setup: Real \gls{ids}}
\label{sub:bona_fide_id}

\begin{figure}
    \hspace*{-1.9cm}
    \centering
    \includegraphics[width=1.3\linewidth]{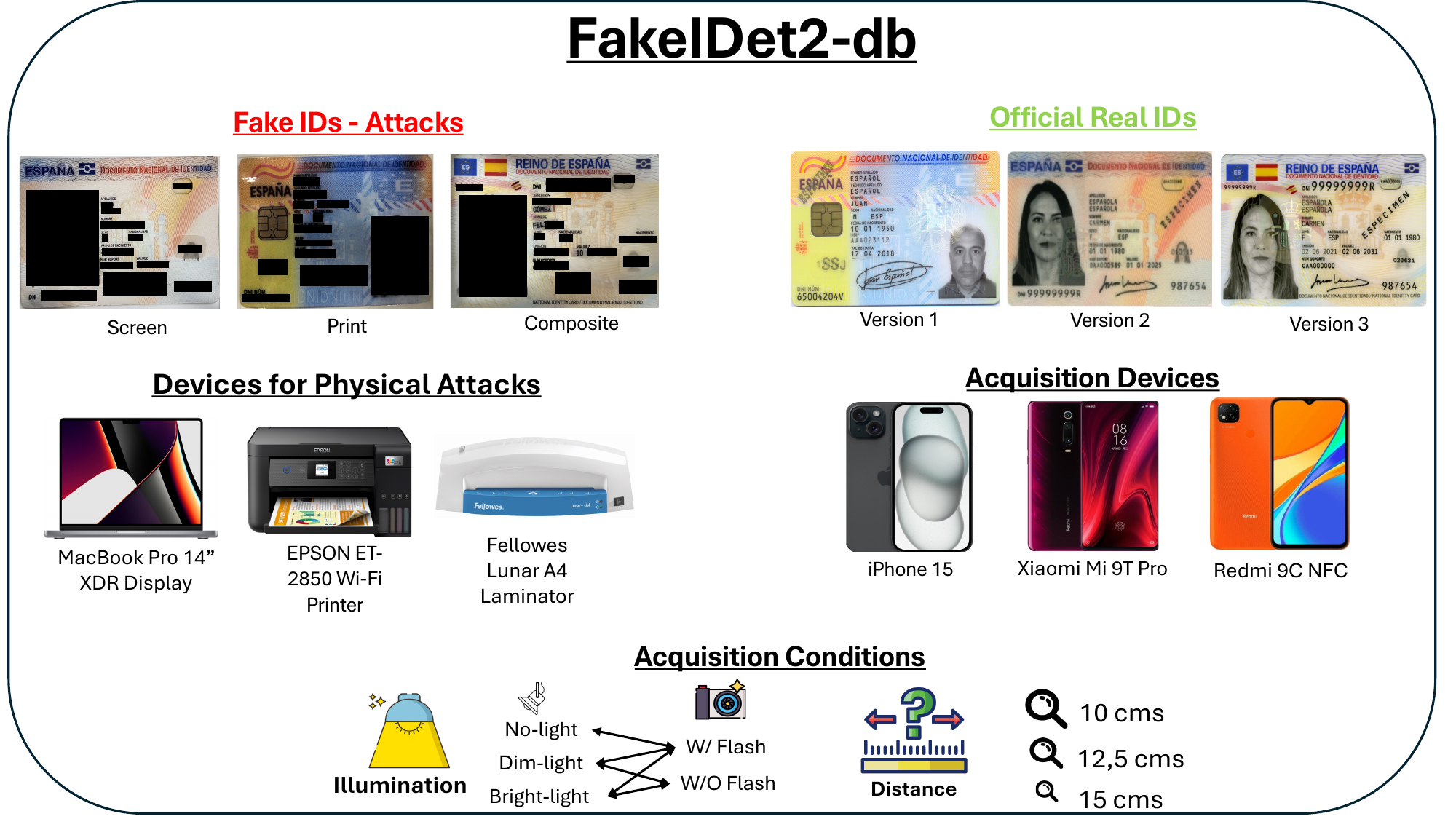}
    \caption{Main characteristics of our proposed public FakeIDet2-db. For the Fake IDs - Attacks, the personal information is redacted in this figure to protect the privacy of the subjects. Official Real IDs are based on template examples.}
    \label{fig:acqu_comp_db}
\end{figure}

Fig.~\ref{fig:acqu_comp_db} summarizes the main characteristics of our proposed FakeIDet2-db. Regarding the number of unique \gls{ids}, our database comprises 47 different official Spanish \gls{ids}. The appearance of Spanish electronic \gls{ids} has changed over the years, which is a key aspect that has also been considered when capturing this database, including three different versions (see the Official Real IDs in Fig.~\ref{fig:acqu_comp_db}). 
For the acquisition, we used three different smartphone models (Apple iPhone 15, Xiaomi Mi 9T Pro, and Redmi 9C NFC), to cover the range of sensor quality from high-end to low-end, respectively. Their main technical specifications are provided in Table~\ref{tab:tech_specs} where we can see that both iPhone 15 and Mi 9T Pro have main camera sensors with 48MP while the Redmi 9C NFC only has 13MP. The iPhone camera is the best in terms of adaptability to low-light scenarios, given its aperture of $f$/1.6, although the Mi 9T Pro is close with an aperture of $f$/1.75. However, the Redmi 9C NFC struggles in these kind of low-light scenarios, since its aperture is greater, $f$/2.2. Regarding image resolution, the iPhone 15 is able to shoot ProRes images up to 8K resolution, but for our acquisition process, we set it to 4K, since both the Mi 9T Pro and the Redmi 9C NFC can only shoot images at that resolution. The photos of \gls{ids} were taken using a 4:3 aspect ratio. Three different heights were considered with respect to the document's vertical: 10 cms, 12.5 cms, and 15 cms, on a 3D printed adjustable platform that we developed specifically for this research. This platform enabled acquiring \gls{ids} precisely, similar to \cite{mudgalgundurao2022pixel}, where they captured \gls{ids} using a mobile application that provided a template in the camera preview to match the \gls{id} borders. 

To provide realistic variability, we also considered different light conditions in terms of room illumination: \textit{i)} \textit{no-light}, with the camera flash activated, \textit{ii)} \textit{dim-light}, with images taken with/without the camera flash activated, and \textit{iii)} \textit{bright-light}, with images taken with/without the camera flash activated. 

Therefore, given the variability factors mentioned above, for a unique \gls{id}, we obtained 45 images (i.e., 3 smartphones $\times$ 3 heights $\times$ 5 light conditions) covering different conditions, generating a total of 2,115 (i.e., 47 $\times$ 45) images of real \gls{ids}.

\begin{table}[t]
    \centering
    \begin{tabular}{lccc}
        \toprule
        \textbf{Feature} & iPhone 15 & Mi 9T Pro & Redmi 9C NFC \\
         \midrule
         \textbf{\makecell{Main Sensor}} & 48 MP (1/1.28") & 48 MP (1/2") & 13 MP (1/3.1") \\
         \textbf{Aperture} & $f$/1.6 & $f$/1.75 & $f$/2.2  \\
         \textbf{Resolution} & 4,032$\times$3,024 & 4,032$\times$3,024 & 4,032$\times$3,024 \\
         \textbf{ISP / SoC} & Apple A16 & Snapdragon 855 & Helio G35 \\
         \textbf{Pixel Size} & \makecell{1.22 $\mu$m/2.44 $\mu$m} & \makecell{0.8 $\mu$m/1.6 $\mu$m} & 1.12 $\mu$m \\
         \bottomrule
    \end{tabular}
    \caption{Technical specifications of the smartphones considered in FakeIDet2-db.}
    \label{tab:tech_specs}
\end{table}

\subsection{Acquisition Setup: Attacks}
\label{sub:attacks_id}
Fig. \ref{fig:id_examples} shows visual examples of physical attacks (print, screen, and physical composite) available in our proposed FakeIDet2-db. Note that personal information is redacted in the figure of the article to protect the privacy of the subject. 

To manufacture the print attacks, we used an HP ScanJet 8270 scanner at 600 \gls{dpi}, creating a digital copy of the real \gls{id}. After that, we printed the fake \gls{ids} using an EPSON ET-2850 Wi-Fi printer and laminated them to improve the realism using a Fellowes Lunar A4 thermal laminator, providing print attacks with the characteristic glossy texture of real IDs. The acquisition process was the same as for real \gls{ids}, yielding a total of 2,115 images for print attacks. 

For screen attacks, we used a MacBook Pro 14" XDR screen, where the 2,115 images of real documents were displayed in full-screen and covering the whole smartphone camera preview. To increase variability between smartphone devices, each image was displayed and then captured using a randomly selected smartphone, ensuring that approximately one third of the images were taken with each device. 

Regarding physical composite attacks, we cut sensitive sections from print attacks and placed them in the sensitive areas of the real \gls{ids}. Regarding privacy considerations, we only cut the date of birth, the name, and both surnames from the \gls{id} owner along with partial sections of the \gls{id}'s number. To provide even more variability, some physical composite attacks were created using cut-outs of several print attacks. The cutouts of sensitive sections from a print attack were never displayed at the same time, further protecting the sensitive information of the \gls{id} owner. Additionally, we did not use the \gls{id} owner's portrait image, as it is sensitive information that cannot be shared under any circumstances. After that, we captured composite attacks following the same procedure as for real \gls{ids}, described in Sec.~\ref{sub:bona_fide_id}, balancing the number of images of real and all types of attack across the database to 2,115.

Gathering all images from real and fake \glspl{id}, our database contains 8,460 total \glspl{id} images. From them, we selected 1,000 images, balancing them across different types of attacks and bona fide data (250 samples per class), different capture devices, and different illumination and height conditions. This subsampling aimed to avoid including images of real and fake IDs under all acquisition conditions, ensuring that fake \gls{id} detectors do not rely on superficial cues, such as illumination patterns, to distinguish between real and fake \glspl{id}, which could lead to overfitting.

\subsection{Post-Processing: Anonymization and Patch Extraction}
\label{sub:fromid2patch}

\begin{figure}[t]
    \centering
    \begin{subfigure}{0.29\linewidth}
        \includegraphics[width=\textwidth]{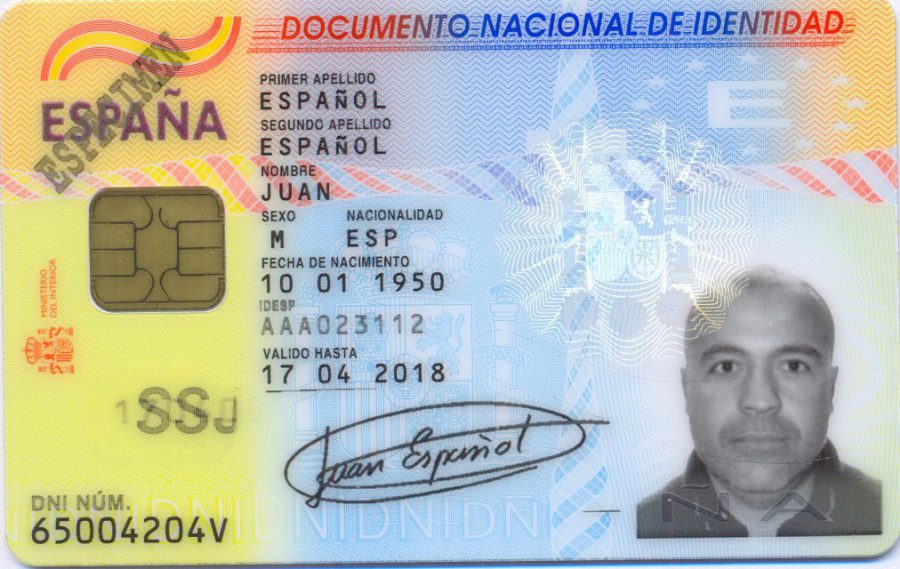}
        \caption{\textit{Non-Anonymized} \gls{id}}
        \label{fig:semi-anon}
    \end{subfigure}
    \hfill
    \begin{subfigure}{0.29\linewidth}
        \includegraphics[width=\textwidth]{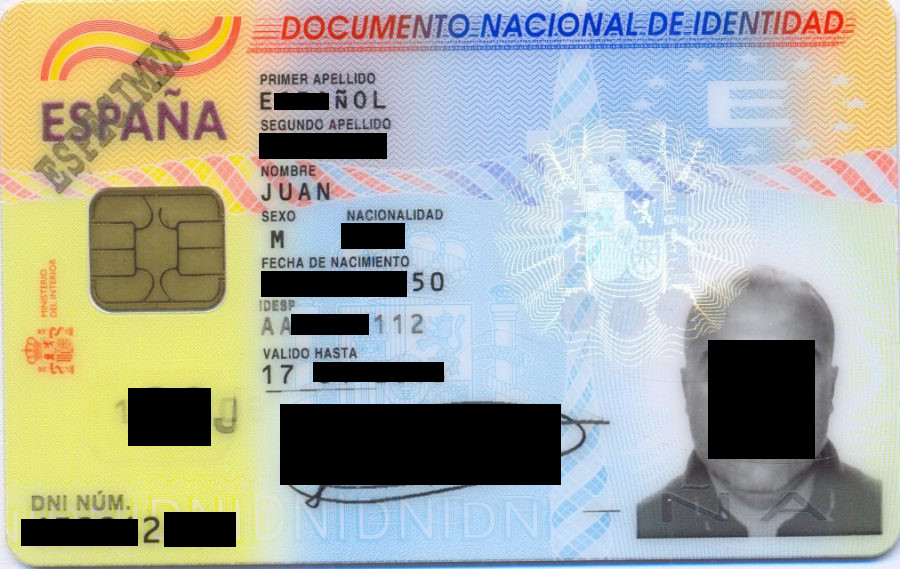}
        \caption{\textit{Pseudo-Anonymized} \gls{id}}
        \label{fig:pseudo-anon}
    \end{subfigure}
        \hfill
        \begin{subfigure}{0.29\linewidth}
        \includegraphics[width=\textwidth]{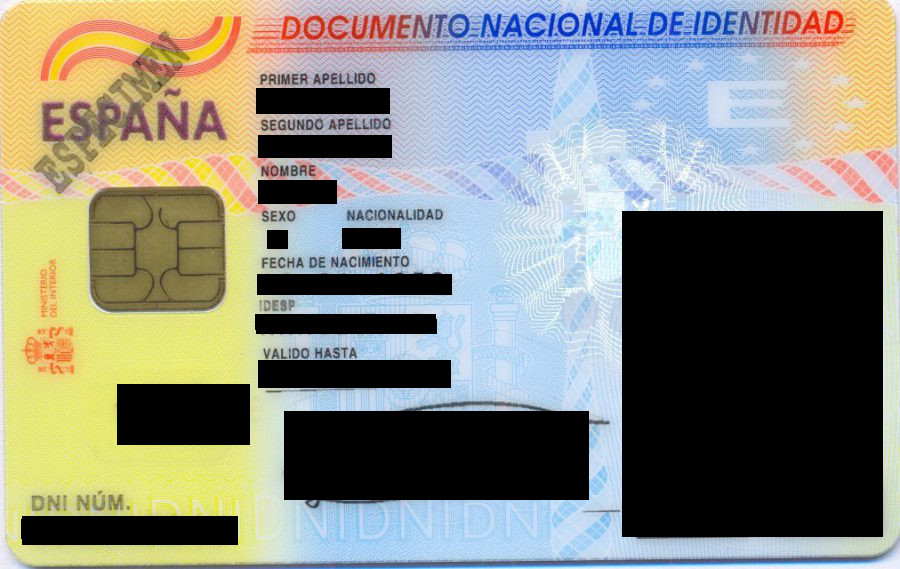}
        \caption{\textit{Fully-Anonymized} \gls{id}}
        \label{fig:fully-anon}
    \end{subfigure}
    \caption{Examples of the different anonymization levels considered in the present study.}
        \label{fig:anon_non-anon}
\end{figure}

Fig.~\ref{fig:anon_non-anon} provides graphical examples of the different anonymization configurations explored in the present study to facilitate collaboration between \gls{id} Holders and AI Researchers in real-world fake \gls{id} detection scenarios. The criteria followed to remove sensitive information at each anonymization configuration are the following:
\begin{enumerate}
    \item \textit{Non-Anonymized}: no sensitive sections are anonymized, so all information is available to detect real/fake IDs.
    \item \textit{Pseudo-Anonymized}: partial sections with sensitive information of real/fake IDs is displayed. This kind of anonymization is consistent along all images from the same \gls{id}, so no reconstruction is allowed.
    \item \textit{Fully-Anonymized}: all sensitive sections of real/fake IDs are completely anonymized.
\end{enumerate}

Referring to \textit{pseudo-} and \textit{fully-}anonymized configurations, we used EasyOCR\footnote{\url{https://github.com/JaidedAI/EasyOCR}} and the GNU Image Manipulator Program (GIMP) to cover the sensitive regions with pitch black rectangles (with the color code (0,0,0) in the RGB spectrum). Since we had two levels of anonymization applied to the original, non-anonymized images, we obtained 2,000 images of real and fake \gls{ids}. Given that we had several images taken from the same \gls{id}, we took special care to cover the same partial sensitive sections across all images in the pseudo-anonymized configuration. Once all \gls{ids} were anonymized, we processed them and extracted patches using PyTorch's unfold method, specifying the same step and stride size as the patch size to avoid overlapping. Patches with a pitch black content greater than 80\% were discarded and the remaining were kept with a sampling probability of $p=0.9$ to further prevent reconstruction. After that, for each \gls{id} image, we created a folder where the extracted patches were stored using random numbers of six figures as the filename of the patch (e.g., 762542.jpeg), so no spatial layout could be inferred.

An important point is the number of patches available with respect to anonymization and patch size configurations, which are depicted in Table~\ref{tab:patches_class_anon_confs} for real \gls{ids} and the different attacks considered. For example, the Full-Anon, $128\times128$ configuration shows a noticeable drop in terms of the number of patches compared to the Non-Anon, $64\times64$ configuration, which has almost ten times fewer patches. As we plan to release FakeIDet2-db, the pseudo- and fully-anonymized \gls{id} configurations at 128$\times$128 and 64$\times$64 patch sizes will be available for privacy reasons, as proved in the results section (Sec. \ref{sec:exp_results}) that these privacy-aware scenarios are very useful to facilitate collaborations between \gls{id} Holders and AI Researchers. If we consider the number of patches from the said configurations, a total of 922,057 patches extracted from 2,000 images will be available to researchers to train and benchmark their proposed fake \gls{id} detectors. 

\begin{table}[t]
    \centering
    \begin{tabular}{cccccc}
        \toprule
        & & \textbf{Real} & \textbf{Print} & \textbf{Screen} & \textbf{Composite} \\
        \midrule
        \textbf{\makecell{Patch \\ Size}} & \textbf{\makecell{Anon. \\ Level}} & & & & \\
        \midrule
        & Non-Anon & 56,017 & 52,976 & 55,927 & 3,810 \\
        $128\times128$ & Pseudo-Anon & 32,875 & 28,432 & 32,170 & 1,805 \\
        & Full-Anon & 21,896 & 20,363 & 21,474 & 1,216 \\
        \midrule
        & Non-Anon & 222,622 & 211,396 & 221,610 & 7,931 \\
        $64\times64$ & Pseudo-Anon & 152,612 & 132,920 & 148,599 & 6,297 \\
        & Full-Anon & 110,162 & 101,178 & 107,104 & 2,954  \\
        \bottomrule
    \end{tabular}
    \caption{Number of patches per real/fake \gls{id}, patch size, and anonymization level considered in our proposed FakeIDet2-db.}
    \label{tab:patches_class_anon_confs}
\end{table}

\begin{figure}[!]
    \centering
    \includegraphics[width=0.91\linewidth]{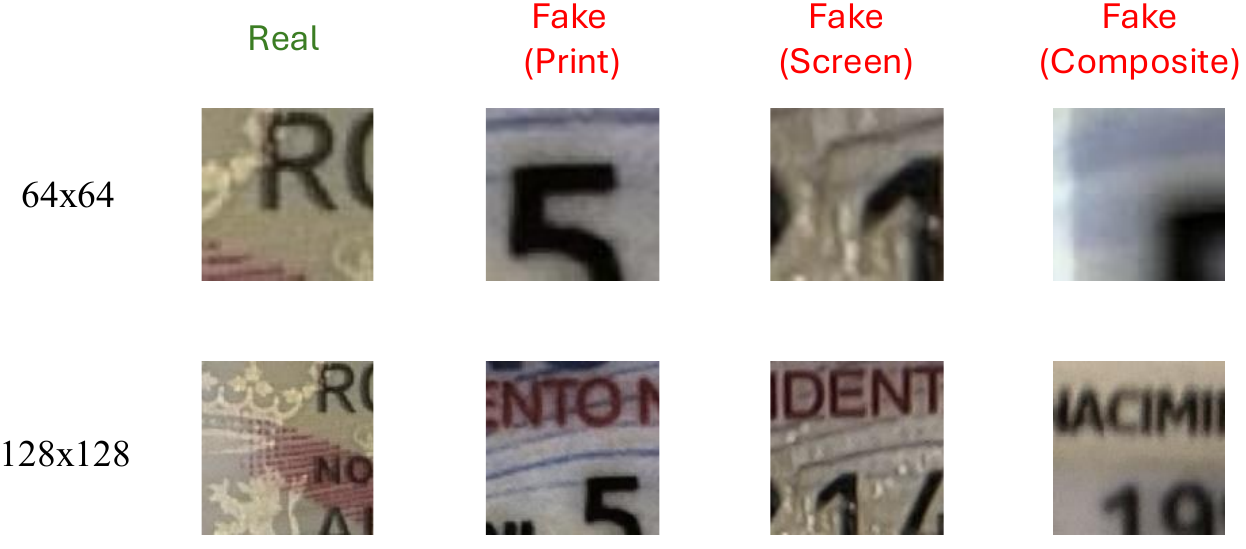}
    \caption{Graphical samples of real/fake patches extracted from different \gls{ids}. Real patches (first column, green color), are shown along the different types of attacks in the proposed database (second, third, and fourth columns, red color), at both 128$\times$128 and 64$\times$64 sizes.}
    \label{fig:patch_example}
\end{figure}

For completeness, Fig.~\ref{fig:patch_example} shows different examples of real/fake patches extracted from different \gls{ids}. One can see that the 128$\times$128 patch size contains much more information compared to their 64$\times$64 counterparts. Some small traces can be observed in the different attacks. For example, composite attacks might show the change in color and texture between the background and the altered section in the foreground. However, patches from screen attacks might show subtle traces of the characteristic Moiré patterns.

\section{Proposed Method for Fake ID Detection: FakeIDet2}
\label{sec:proposed_method}

\begin{figure}
    \centering
    \includegraphics[width=0.66\linewidth]{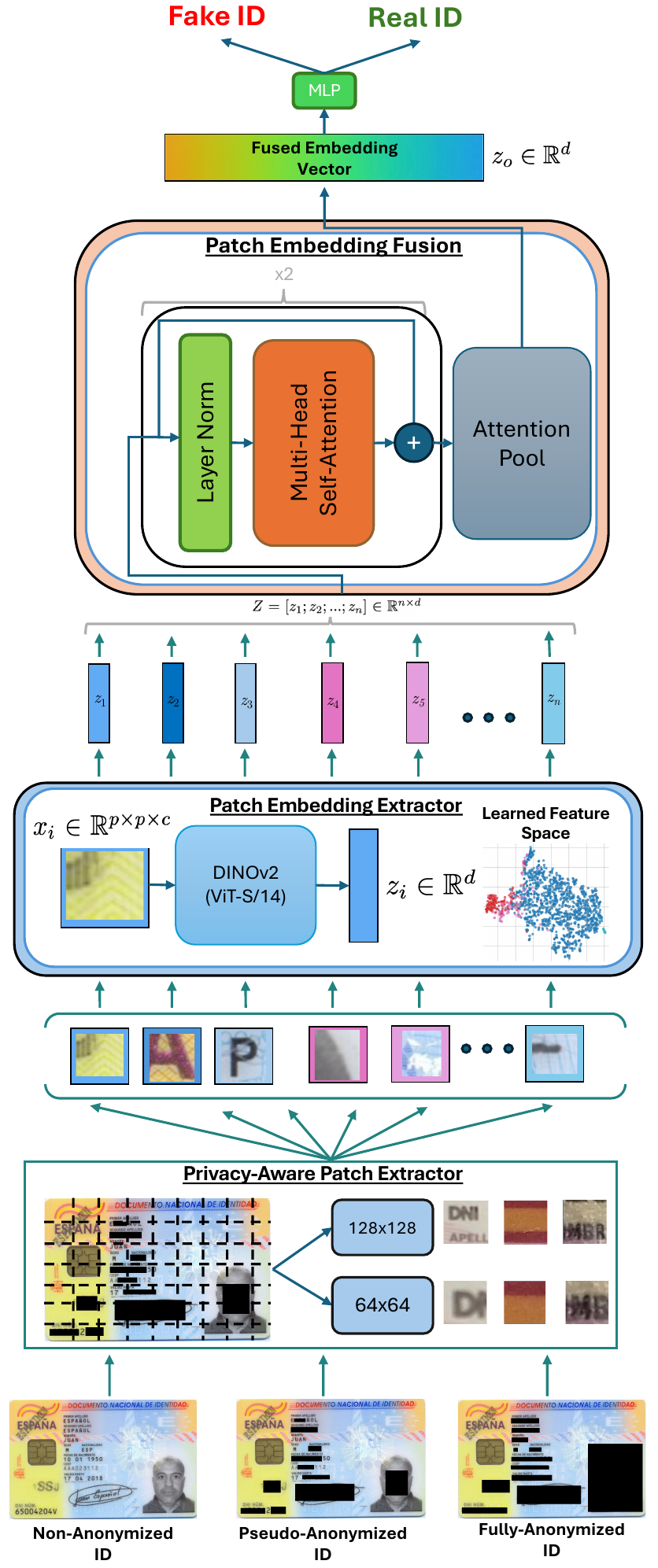}
    \caption{General diagram of our proposed fake \gls{id} detector, FakeIDet2. Our proposal has three different stages: \textit{i)} the entire \gls{id} is forwarded to the Privacy-Aware Patch Extractor, extracting patches of fixed size (128$\times$128 or 64$\times$64); \textit{ii)} those patches are forwarded into the Patch Embedding Extractor, which yields an embedding per patch; and \textit{iii)} finally, those embeddings are used at the Patch Embedding Fusion module to generate a single embedding which is used by an \gls{mlp}, providing a final score that is thresholded to obtain the final decision: real or fake ID.}
    \label{fig:fakeidet-v2}
\end{figure}

Our proposed fake \gls{id} detection method, FakeIDet2, proposes a privacy-aware, patch-level representation fusion \cite{fierrez18fusion}. Fig.~\ref{fig:fakeidet-v2} provides a graphical representation of FakeIDet2. This comprises three main modules: the Privacy-Aware Patch Extractor, the Patch Embedding Extractor, and the Patch Embedding Fusion. The Privacy-Aware Patch Extractor module receives a complete \gls{id}, considering any anonymization level presented in Sec.~\ref{sub:fromid2patch}. This first module divides the entire \gls{id} into patches, removing sensitive information if applicable (i.e., the regions of the \gls{id} that are blacked out). After that, those patches are forwarded to the Patch Embedding Extractor, which uses a fine-tuned DINOv2 as a feature extractor to learn discriminative features at the patch level of each type of \gls{pai}, providing an embedding for each patch. Then, those embeddings are treated as tokens, and passed into a \gls{mhsa} layer, which ponders the embeddings values depending on their importance in the sequence. After that, an attention pool layer selects discriminative features from the sequence, obtaining a single embedding, which is forwarded to a \gls{mlp} that yields a final score per \gls{id}, indicating whether it is real or fake. Next, we describe the technical details of the \textit{learnable} modules of FakeIDet2.

\subsection{Patch Embedding Extractor}
\label{sub:patch_rep_learning}

To effectively learn features at patch level, we go deep into representation learning, aiming to obtain compressed, meaningful representations from high-dimensional data into the embedding space. To do so, we treat patches coming from both real and fake \gls{ids} as four different classes (i.e., 3 \gls{pai}s + real) and use DINOv2's (ViT-S/14) \cite{oquab2023dinov2} backbone with its weights frozen. Using DINOv2 for extracting patch embeddings seems appropriate, since the goal of its original training procedure was to match the embedding distribution between images where the full context is given and random patches extracted from the same image. To take advantage of these features, we attach a final layer to DINOv2 embeddings; hence, our Patch Embedding Extractor is defined as an encoder function that maps an input patch $x_i \in \mathbb{R}^{p \times p \times c}$, where $p$ is the patch size and $c$ is the channel dimension (i.e., $c=3$ in the RGB spectrum), to its corresponding compressed representation $z_i \in \mathbb{R}^d$, where $d$ is the dimension of the embedded space. 

We explore three softmax-based loss functions \cite{2022_PR_SetMargin_Morales}: CosFace \cite{wang2018cosface}, ArcFace \cite{deng2019arcface}, and AdaFace \cite{kim2022adaface}, to assess which one best fits our problem to learn discriminative representations. The generic formulation for the softmax loss function is defined in Eq. \ref{eq:softmax_based}, where $C$ is the number of classes, $f(\theta_{y_i})$ is a function that computes the alignment between the embedding obtained from the input image and the class prototype of its label $y_i$ and $s$ is a scaling factor that modulates the alignment value. Hence, the objective for an embedding of one class is to be as close as possible to other embeddings from the same class (better aligned) while being apart from embeddings of other classes by an imposed distance, called margin (worst aligned). The main differences between these loss functions reside in the way the margin is computed for the alignment function, $f(\theta_{y_i})$. For example, CosFace, defined in Eq. \ref{eq:cos_margin_func}, computes the loss by adding the margin $m$ to the cosine similarity, while ArcFace, defined in Eq. \ref{eq:arc_margin_func}, computes the loss by adding an angular margin $m$ to the cosine similarity. In the case of AdaFace, defined in Eq. \ref{eq:ada_margin_func}, the margin is adaptively calculated, depending on the quality of the image \cite{2022_CSUR_FaceQsurvey_Torsten}, which is calculated as the embedding norm obtained from the input image normalized on batch statistics for better stability and is denoted by $||\widehat{z_i}||$. 

\begin{equation}
    \mathcal{L} = \log\left( \frac{\text{exp}(f(\theta_{y_i}))}{\text{exp}(f(\theta_{y_i})) + \sum_{j \neq y_i}^{n} \text{exp}(s\cos\theta_j)} \right)
    \label{eq:softmax_based}
\end{equation}

\begin{equation}
    f(\theta_{y_i})_{\text{CosFace}} = \begin{cases}
        s\cos(\theta_j - m) & j = y_i \\
        s\cos(\theta_j) & j \neq y_i
    \end{cases}
    \label{eq:cos_margin_func}
\end{equation}

\begin{equation}
    f(\theta_{y_i})_{\text{ArcFace}} = \begin{cases}
        s\cos(\theta_j + m) & j = y_i \\
        s\cos(\theta_j) & j \neq y_i
    \end{cases}
    \label{eq:arc_margin_func}
\end{equation}

\begin{equation}
    f(\theta_{y_i})_{\text{AdaFace}} = \begin{cases}
        s\cos(\theta_j + (-m \cdot ||\widehat{z_i}||)) - (m \cdot ||\widehat{z_i}|| + m) & j = y_i \\
        s\cos(\theta_j) & j \neq y_i
    \end{cases}
    \label{eq:ada_margin_func}
\end{equation}

An issue we encountered when training the Patch Embedding Extractor was that softmax-based loss functions did not account for the severe class imbalance in the physical composite attacks (see Table \ref{tab:patches_class_anon_confs}). To address this limitation, we propose to apply dynamic class weights to these loss functions that evolve as training progresses \cite{fierrez18fusion}. The initial weights $w_0 \in \mathbb{R}^C$, where $C$ is the number of classes, are calculated as the normalized inverse of the frequency for each class using Eq.~\ref{eq:class_weights_computation}, where $N$ is the total number of patches in the database and $N_i$ is the total number of patches belonging to the $i$-th class.

\begin{equation}
    w_0 = \begin{bmatrix} \frac{N}{N_1} \\ \frac{N}{N_2} \\ ... \\ \frac{N}{N_C}\end{bmatrix} \times \sum_{i=1}^{C} \frac{C}{\frac{N}{N_{i}}}
    \label{eq:class_weights_computation}
\end{equation}

After that, the class weights are updated at each epoch until they all converge to 1, hence progressively giving the same weight to all classes. The update rule is presented in Eq. \ref{eq:class_weights_upd_rule}, where $w_t \in \mathbb{R}^C$ is a vector that stores the weights of the class at the time step $t$, $\mathbf{1} \in \mathbb{R}^C$ is a vector full of ones, $\lambda_t \in [0,1]$ is a pondering factor that linearly decays from 1 to 0 as training progresses, and $e$ is the total number of time steps involved in the training procedure and $\lambda_e=0$. Intuitively, at the beginning of the training, the class weights will be close to $w_0$, giving more importance to the underrepresented class, but, as time goes by, the class weights will be closer to 1, giving all classes the same importance at the end of the training.

\begin{equation}
    w_{t} = (1-\lambda_{t})\mathbf{1} + \lambda_{t}w_0, \quad
    \lambda_t \leftarrow \lambda_0 + (\lambda_e - \lambda_0)\times\left(\frac{t}{e}\right)
    \label{eq:class_weights_upd_rule}
\end{equation}

These class weights are then used to modulate the gradient signal that the margin loss function produces, giving higher values when the embeddings of the underrepresented class are close to the embeddings from another class. The modified version of this loss is defined in Eq. \ref{eq:weighted_adaface}, where $\mathcal{L}$ is the loss function defined in Eq. \ref{eq:softmax_based}, $w_{t, y_i}$ is the class weight associated with the label $y$ of the $i$-th element of the batch at the time $t$. 

\begin{equation}
    \mathcal{L}_{\text{weighted}} = -w_{t, y_i} \cdot \mathcal{L}
    \label{eq:weighted_adaface}
\end{equation}

\subsection{Patch Embedding Fusion}
\label{sub:patch_rep_fusion}
To effectively fuse all patch embeddings of a given \gls{id}, we use a Multi-Head Self-Attention (MHSA) layer \cite{alcala25attzoom}. This self-attention mechanism was introduced to solve problems related to long-range dependency issues of the model, which are typical in \gls{rnn} models, which process information sequentially. However, in the self-attention mechanism, all elements of the input sequence are processed at the same time, enabling modeling of dependencies from distant tokens. 
The Self-Attention layer expects as input a sequence of embeddings (or tokens) $X \in \mathbb{R}^{n \times d}$, where $n$ is denoted as the sequence length and $d$ the embedding dimension. The sequence is processed according to the Self-Attention mechanism, defined in Eq. \ref{eq:self_attn}, where $Q$ is the query, $K$ is the key, $V$ is the value, and $d_k$ is the projected embedding dimension for the query and the key. The query, key, and value matrices are single linear projections of $X$, which are calculated as $Q=XW_Q$, $K=XW_K$ and $V=XW_V$, respectively, where $W_Q \in \mathbb{R}^{d \times d_k}$, $W_K \in \mathbb{R}^{d \times d_k}$ and $W_V \in \mathbb{R}^{d \times d_v}$, where $d_v$ is the projected dimension of the value. Hence, the sequence obtained after the self-attention mechanism is a matrix $X' \in \mathbb{R}^{n \times d}$ where each token value has been modified depending on their importance in the sequence, compared to their peers.  

\begin{equation}
    \mathrm{Attention}(Q, K, V) = \mathrm{softmax}\left( \frac{QK^\top}{\sqrt{d_k}} \right)V
    \label{eq:self_attn}
\end{equation}

Furthermore, the attention mechanism allows queries, keys, and values to be linearly projected $h$ times using different learned projections of dimensions $d_k$ and $d_v$. To achieve this, $d_k$ and $d_v$ are typically set to $d / h$, where $d$ is the dimension of the model. Each of these computations defines an attention head (Eq.~\ref{eq:attn}). The outputs of the $h$ attention heads are concatenated and projected back to the original input dimension using a final linear transformation $W^O \in \mathbb{R}^{h d_v \times d}$ (Eq.~\ref{eq:msha}).

\begin{equation}
\text{head}_i = \mathrm{Attention}(Q_i, K_i, V_i)
\label{eq:attn}
\end{equation}

\begin{equation}
\mathrm{MultiHead}(Q, K, V) = [\text{head}_1, \ldots, \text{head}_h]W^O
\label{eq:msha}
\end{equation}

Since the Patch Embedding Extractor produces an embedding $z_i$ for each patch $x_i$, we treat the set of embeddings that come from a single \gls{id} as a sequence $Z = [z_1; z_2; ...; z_n] \in \mathbb{R}^{n \times d}$, where $n$ is the number of patches from a single \gls{id} and $d$ is the dimension of the embedding obtained from the Patch Embedding Extractor. These embeddings are then normalized through layer normalization \cite{ba2016layer} and are transmitted to the \gls{mhsa} layer that produces a feature sequence $Z' = [z_1'; z_2'; ...; z_n'] \in \mathbb{R}^{n \times d}$ where the embedding values $z_i'$ have been modified according to their importance in the input sequence. Furthermore, we leverage residual connections to enable a better gradient flow in the training procedure, since two layers of \gls{mhsa} and layer normalization are stacked in this module, where the first \gls{mhsa} layer has $h=8$ attention heads and the second layer has $h=4$ attention heads. 

Using the \gls{mhsa} mechanism is beneficial for our proposed FakeIDet2 as: \textit{i)} it leverages the patch embedding space learned by the Patch Embedding Extractor, and \textit{ii)} it can detect elements of the sequence that may be anomalous, which is analogous to detecting tampered sections of a composite attack. Then, the feature sequence is forwarded into a temporal attention pool module, similar to \cite{attn_pool}, generating a single embedding $z_o \in \mathbb{R}^d$. Finally, we use $z_o$ as an input to an \gls{mlp}, comprising a single layer with sigmoid activation, obtaining a final score per \gls{id}.

\section{Experimental Setup}
\label{sec:experiments}

In order to define a standard, reproducible benchmark that allows us to advance properly in this research field, in this section, we define our experimental setup. First, Sec. \ref{sub:train_settings} describes the specific hyperparameters of our proposed FakeIDet2 method. Then, in Sec. \ref{sub:exp_proto} we describe the experimental protocol and the different scenarios of interest in the analysis. Finally, Sec. \ref{sub:ev_metrics} describes the evaluation metrics. In general, we use the \gls{eer} metric as a performance indicator to obtain the best patch and anonymization configuration. For all training procedures and experiments, we used only one NVIDIA RTX 3080 with 10GB of VRAM.

%configurations for the training procedures of the models, detailing such things as learning rate, number of layers, different learning rate schedulers, etc. Furthermore, we also include different scenarios of interest for this kind of systems, such as training leaving one attack or sensor out from the training data. At last, we also propose to use other databases to test our model performance in out-of-distribution data, specially when detecting attacks that has been captured with different sensors, or even generated synthetically. As a general note, we use the \gls{eer} metric as a performance indicator to obtain the best patch and anonymization configuration. For all training procedures and experiments, we used only one NVIDIA RTX 3080 with 10GB of VRAM.

\subsection{FakeIDet2: Hyperparameters}
\label{sub:train_settings}
Regarding the Patch Embedding Extractor, for any patch size, the input is resized to $224\times224$, which is the input shape of DINOv2, with an embedding layer of dimension $d=128$. We use Adam optimizer with weight decay, an initial learning rate $\alpha_0 = 0.00125$, and exponential decays for the momentum estimators of $\beta_1 = 0.9$ and $\beta_2 = 0.999$, respectively, and a weight decay $\delta=0.0001$. The modified versions of the softmax-based losses presented in Sec.~\ref{sub:patch_rep_learning} are defined with a margin of $m=0.4$, which are the optimal experimental values that perform well for low-quality datasets without compromising the capacity in high-quality data, and a scaling factor of $s=64$, as it provides a strong signal from both samples that are close and far from the decision boundary, creating better separability in the learned representations according to \cite{kim2022adaface}. All models are trained for 70 epochs, with an early stopping condition that ends the training if the loss in the validation does not decrease for 5 consecutive epochs. Furthermore, we use a cosine annealing decay scheduler for the learning rate, with values decreasing from $\alpha_0 = 1.25e-3$ to $\alpha_e = 1.25e-4$, and a batch size of 256. Regarding data transformations, we opt to apply random Gaussian blur with a kernel size $k=3$ and color jitter with a brightness of 0.2, a hue of 0.05, a contrast of 0.25, and a saturation of 0.2. These data transformations are applied in training time with a probability of $p=0.2$. Finally, the pixel values are normalized according to ImageNet mean and standard deviation statistics. 

For the Patch Embedding Fusion, in particular the \gls{mhsa} layers, we define a context length of 384 for both patch configurations, which is enough so that no patch from an \gls{id} is left out when performing a forward pass. Furthermore, we use masked-\gls{mhsa} to indicate which embeddings should be attended, so that in case the number of patches from a single \gls{id} is fewer than the context length, the introduced padded embeddings are ignored. We use \gls{bce} loss with class weights, since there is a class ratio of 3 to 1 between the fake and the real data. For all of our experiments, we choose Adam as optimizer, using an initial learning rate $\alpha_0 = 1.25e-4$ and exponential decays for the momentum estimators of $\beta_1 = 0.9$ and $\beta_2 = 0.999$, respectively, with 1 epoch of learning rate warm-up. Furthermore, the learning rate followed a cosine annealing decay from $\alpha_0 = 1.25e-4$ to $\alpha_e = 1.25e-5$. 

All configurations in terms of anonymization levels and patch sizes are trained for 10 epochs with a batch size of 4, selecting the best performing model for each training procedure according to the lowest validation loss achieved. 
 
\subsection{Experimental Protocol}
\label{sub:exp_proto}
In order to assess the feasibility of our proposed FakeIDet2 detector and privacy-aware scenarios to facilitate collaboration between \gls{id} Holders and AI Researchers, we propose an experimental protocol that aims to: \textit{i)} evaluate FakeIDet2 performance regarding patch size and anonymization configurations in our new FakeIDet2-db, comparing the results with the traditional case in the literature, i.e., feeding the fake detectors with the whole non-anonymized \gls{id}, and \textit{ii)} explore domain adaptation, by evaluating FakeIDet2 on popular databases in the topic of fake \gls{id} detection.

First, in all experiments that involve any kind of training procedure, the database is divided into a development set (80\% of the real and fake IDs) and a final evaluation set (the remaining 20\%) for all training scenarios. This split simulates a realistic scenario in which the evaluation involves unseen identities. In addition, for our FakeIDet2-db, the first version of Spanish \gls{id} is included only in the final evaluation set, not in the development set, to analyze the generalizability of FakeIDet2 to other \gls{id} templates. 

Regarding the experiments included in Sec. \ref{sec:exp_results}, we first validate the technical details of our proposed FakeIDet2 in terms of performance and privacy. Sec.~\ref{sub:select_loss_function} explores the different softmax-based loss functions for representation learning, which will determine the best loss function for the Patch Embedding Extractor module. After that, an exploration in terms of the optimal patch size is conducted in Sec.~\ref{sub:ps_vs_performance}, to test whether more patches at lower sizes (i.e., $64\times64$) yield better performance compared to fewer patches of greater size (i.e., $128\times128$). Additionally, we also compare the performance of our proposed FakeIDet2 in terms of anonymization levels in Sec.~\ref{sub:anon_vs_performance}, which also goes deep into interesting scenarios such as how well a model trained on fully-anonymized data performs on non-anonymized data. Finally, for completeness, we compare FakeIDet2 with our previous FakeIDet \cite{munoz2025exploring}. From this set of experiments, we will select the best configuration of FakeIDet2 in terms of privacy versus performance.

Then, after validating the technical contributions of FakeIDet2, we analyze its generalization capabilities against unseen scenarios, including leave-one-out experiments for two interesting scenarios: novel attack detection and out-of-distribution sensors. In this case, we train FakeIDet2 with the optimal configuration, leaving images belonging to one \gls{pai} or images taken from one sensor out of the training data, testing the performance of FakeIDet2 when exposed to unseen attacks, or images taken from acquisition devices with sensors that were not used during training. The analyses of these scenarios are available in Sec. \ref{sub:leave_one_attack_out} and Sec. \ref{sub:leave_one_sensor_out}, respectively.

Finally, we evaluate in Sec.~\ref{sec:x_db_scenario} the best configuration of FakeIDet2 in a cross-database scenario, providing a standard public benchmark for the research community. This scenario aims to evaluate our proposed model using unseen databases, where the IDs are from different countries, including different templates, and are acquired under different conditions in terms of illumination and acquisition devices. In addition, these databases contain different \gls{pais} compared to our proposed database. With this goal in mind, we select three different databases for performance evaluation, DLC-2021 \cite{polevoy2022document}, KID34K \cite{kid34k} and Benalcazar \textit{et al.} \cite{synth_id_card_db}. These databases are selected for the following reasons: \textit{i)} each database has \gls{id} images from three different regions of the world (i.e., Europe, Asia and South America); hence the evaluation is conducted on different distributions in terms of \gls{id} templates with different alphabets, and \textit{ii)} we contemplate physical attacks as well as synthetic attacks in our evaluation, to assess the performance of FakeIDet2 to synthetic data in the form of fake IDs.

\subsection{Evaluation Metrics}
\label{sub:ev_metrics}
Similar to previous approaches presented in the literature~\cite{tapia2024first}, we adopt the ISO/IEC 30107-3 standard\footnote{\href{https://www.iso.org/standard/79520.html}{https://www.iso.org/standard/79520.html}} for the evaluation of fake \gls{id} detection technology. This includes the use of the \gls{bpcer} and \gls{apcer} metrics. The \gls{bpcer} measures the proportion of bona fide samples (i.e., real IDs) that are incorrectly classified as attacks (i.e., fake IDs), see Eq.~\ref{eq:bpcer}, where $N_\textrm{BF}$ is the number of bona fide samples, $\textrm{RES}_{i}$ is a term that is equal to 1 if the presented sample is incorrectly classified as an attack, and $\tau$ is the operational point, i.e., the threshold used to classify a sample as bona fide if the confidence of the prediction goes below and as an attack otherwise.

\begin{equation}
    \textrm{BPCER}(\tau) = \frac{\sum^{N_\textrm{BF}}_{i=1} \textrm{RES}_i}{N_\textrm{BF}}
    \label{eq:bpcer}
\end{equation}

Similarly, \gls{apcer} quantifies the proportion of attack samples (i.e., fake IDs) that are misclassified as bona fide (i.e., real IDs): 

\begin{equation}
    \textrm{APCER}(\tau) = \frac{1}{N_\textrm{BF}}\sum^{N_\textrm{BF}}_{i=1} 1 - \textrm{RES}_i
    \label{eq:apcer}
\end{equation}

In addition, we consider the Equal Error Rate (EER), a widely used metric in the literature. The EER corresponds to the error rate at the operational threshold $\tau$ where $\textrm{\gls{bpcer}}(\tau) = \textrm{\gls{apcer}}(\tau)$, providing a summary of the system performance in a single value.
\section{Experimental Results}
\label{sec:exp_results}

This section evaluates the performance of our proposed FakeIDet2. The performance results included in all tables are at the ID level, not the patch level, on the final evaluation datasets. This facilitates the comparison of our proposed FakeIDet2 with traditional approaches in the literature, focused on the whole ID level evaluation.

First, we evaluate the technical novelties of our proposal. Sec. \ref{sub:select_loss_function} provides a comprehensive comparison in terms of softmax-based loss functions. The analysis of the optimal patch size configuration and anonymization level is carried out in Sec.~\ref{sub:ps_vs_performance} and Sec.~\ref{sub:anon_vs_performance}, respectively. Finally, for completeness, we compare FakeIDet2 with our previous FakeIDet \cite{munoz2025exploring}. These first experiments are carried out using our new FakeIDet2-db, as described in Sec. \ref{sub:exp_proto}.

Once the preliminary analysis is conducted using FakeIDet2-db, we select the optimal configuration balancing performance and privacy and use it to evaluate FakeIDet2 in two common scenarios for \gls{pad}. The first scenario is introduced in Sec.~\ref{sub:leave_one_attack_out}, which studies the performance of FakeIDet2 to unseen attacks. Hence, we train FakeIDet2 with the optimal configuration leaving one of the \gls{pais} out of the training set. The second scenario, Sec. \ref{sub:leave_one_sensor_out}, analyzes the performance of FakeIDet2 when exposed to images taken with smartphone cameras that have not been seen in training. To do so, we train FakeIDet2 using images from all camera sensors except one. Finally, we evaluate in Sec.~\ref{sec:x_db_scenario} the best configuration of FakeIDet2 in a cross-database scenario, providing a standard public benchmark for the research community.

\subsection{Selecting the Optimal Loss Function Configuration}
\label{sub:select_loss_function}

Given that different softmax-based loss functions are considered in FakeIDet2 training, we perform a comparison in terms of performance using the 128$\times$128, Non-Anonymized configuration. For each loss function, we also compare our proposed dynamic class weights (Dynamic-Weights) described in Sec. \ref{sub:patch_rep_learning} with traditional approaches considered in the literature:
\begin{itemize}
    \item No-Weights: All classes have a fixed weight equal to one.
    \item Static-Weights: Each class has its own weight based on its frequency in the dataset (Eq. \ref{eq:class_weights_computation}), which is kept fixed during training.   
\end{itemize}

Table \ref{tab:lossfn_comp_performance} shows the results for the different loss functions and different types of class weights during learning. We observe that for all loss functions, the proposed Dynamic-Weights achieve better overall results. For example, in the case of CosFace we see that the No-Weights and Static-Weights configurations achieve practically the same results with 4.32\% \gls{eer}, while the Dynamic-Weights improve performance to 3.66\% \gls{eer}. ArcFace shows minor improvements using Dynamic-Weights, with an improvement of 0.34\% \gls{eer} and 1.67\% \gls{eer} with respect to using Static-Weights and No-Weights, respectively. The best performance is obtained with AdaFace using Dynamic-Weights, with 2.01\% \gls{eer}, providing an improvement of 1.65\% \gls{eer} over No-Weights, and 4.30\% \gls{eer} over Static-Weights. 

\begin{table}[t]
\centering
\begin{tabular}{lcccc}
\toprule
\textbf{Model} & \textbf{\makecell{Screen}} & \textbf{\makecell{Print}} & \textbf{\makecell{Composite}} & \textbf{All} \\
\midrule
& & \textit{CosFace} \cite{wang2018cosface} & & \\
\midrule
No-Weights & 3.00 & 1.00 & 6.12 & 4.32 \\
Static-Weights & \textbf{1.02} & 3.89 & 7.17 & 4.32 \\
Dynamic-Weights & 4.00 & \textbf{0.00} & \textbf{4.08} & \textbf{3.66} \\
\midrule
& & \textit{ArcFace} \cite{deng2019arcface} & & \\
\midrule
No-Weights & 6.00 & 6.77 & 10.21 & 7.32 \\
Static-Weights & 5.08 & 5.83 & 8.17 & 5.99\\
Dynamic-Weights & \textbf{1.12} & \textbf{2.89} & \textbf{7.13} & \textbf{5.65} \\
\midrule
& & \textit{AdaFace} \cite{kim2022adaface} & & \\
\midrule
No-Weights & 6.23 & \textbf{0.00} & 4.08 & 3.66 \\
Static-Weights & 6.00 & 7.77 & 5.12 & 6.31\\
Dynamic-Weights & \textbf{1.23} & 0.12 & \textbf{2.09} & \textbf{2.01} \\
\bottomrule
\end{tabular}
\caption{Results in terms of EER (\%) for the different loss functions (CosFace, ArcFace, and AdaFace) and different types of class weights during learning. The proposed Dynamic-Weights allow to deal with class imbalance among the different \gls{pais} patches. Experiments are conducted on the entire evaluation set of FakeIDet2-db, considering the configuration 128$\times$128 Non-Anonymized.}
\label{tab:lossfn_comp_performance}
\end{table}

\begin{figure}[!]
    \centering
    \includegraphics[width=1\linewidth]{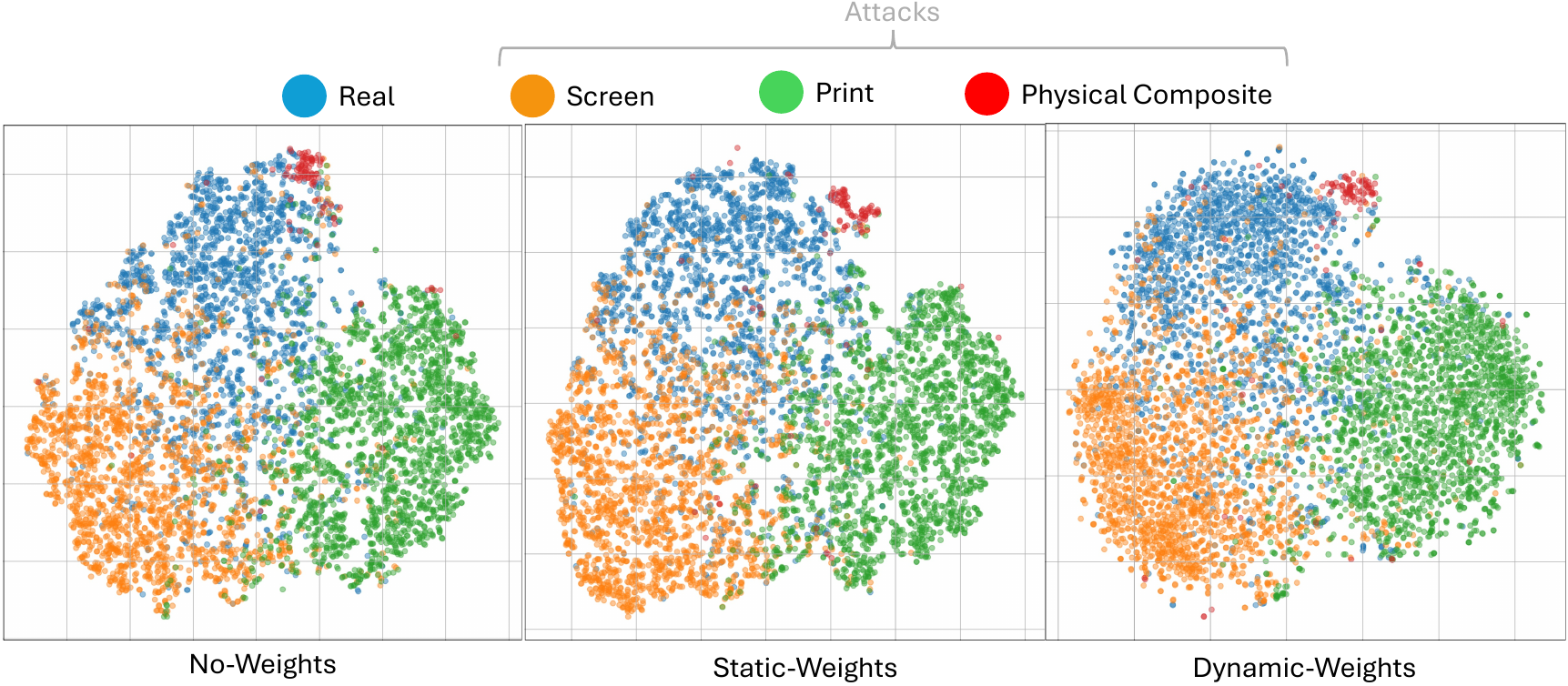}
    \caption{Differences of AdaFace's learned embedding space using No-Weights (left), Static-Weights (middle), and Dynamic-Weights (right) via t-SNE 2D.}
    \label{fig:cos_vs_arc_vs_ada}
\end{figure}

Furthermore, in order to prove that the improvement in performance correlates with how the embedded space is modeled, we plot in Fig.~\ref{fig:cos_vs_arc_vs_ada} the embeddings distributions of AdaFace using the t-SNE projection in 2D. In the figure, we show visual examples when: \textit{i)} No-Weights are applied (left), \textit{ii)} Static-Weights are applied (middle), and \textit{iii)} Dynamic-Weights are applied (right). The data used for this visualization were randomly sampled from the test set, comprising 6,000 patches.

In the No-Weights case, we observe that the physical composite patches (red points) overlap with the real patches (blue points). This is a critical issue, as confusion between these two classes can significantly hinder the detection of physical composite attacks at the document level, indicating that the embeddings are not sufficiently discriminative. 

For Static-Weights, we observe a clearer clustering trend for physical composite patches. However, as expected, keeping the weight values fixed throughout the training biases the model away from properly clustering patches from the other classes (i.e., real, print, and screen), which appear more scattered across the embedding space.

In contrast, with our proposed Dynamic-Weights, we observe a clear separation between physical composite patches and those from the remaining classes. This is enabled by the initial class weights computed at the start of training. Moreover, the clustering of the other classes is more consistent, with less sparsity in the boundaries between clusters and more sparsity within clusters compared to No-Weights and Static-Weights, which indicates a more robust feature extraction. We attribute this behavior to the later training stages using adaptive weights, which guide the model to better cluster patches from the more prevalent classes in the dataset.

The results of this section showcase that our proposed dynamic class weights approach better suits the problematic scenario of heavy class-imbalance. In summary, we adopt the AdaFace loss function with Dynamic-Weights for the rest of our experiments.

\subsection{Patch Size vs. Performance}
\label{sub:ps_vs_performance}

For this section, we consider the Non-Anonymized configuration in order to provide a direct comparison of our proposed FakeIDet2 approach, based on patches, with the current scenario in the literature, based on the entire Non-Anonymized \gls{id}. For completeness, we also include in the comparison our previous FakeIDet detector \cite{munoz2025exploring}, trained using 128$\times$128 and 64$\times$64 patch sizes, similar to FakeIDet2. For the scenario of feeding the fake detector with the whole \gls{id}, we fine-tune the DINOv2 foundation model \cite{oquab2023dinov2} using a final sigmoid layer attached to the DINOv2 output embedding. 

Table \ref{tab:patch_comp_performance} shows a comparison between the performance of the fine-tuned DINOv2 (whole \gls{id}), FakeIDet \cite{munoz2025exploring}, and FakeIDet2 in terms of \gls{eer}. In general, we see that training on the whole \gls{id} does not achieve good results (22.63\% \gls{eer}), in contrast to our proposed patch-wise approach. 

\begin{table}
\centering
\begin{tabular}{lcccc}
\toprule
\textbf{Model} & \textbf{\makecell{Screen}} & \textbf{\makecell{Print}} & \textbf{\makecell{Composite}} & \textbf{All} \\
\midrule
& & \textit{Entire \gls{id}} & & \\
\midrule
DINOv2 \cite{oquab2023dinov2} & 21.85 & 11.35 & 24.68 & 22.63 \\
\midrule
& & \textit{Patch Size: 128$\times$128} & & \\
\midrule
FakeIDet~\cite{munoz2025exploring} & 7.10 & \textbf{0.00} & 54.08 & 25.58 \\
\bf{FakeIDet2} & \textbf{1.23} & 0.12 & \textbf{2.09} & \textbf{2.01} \\
\midrule
& & \textit{Patch Size: 64$\times$64} & & \\
\midrule
FakeIDet~\cite{munoz2025exploring} & 9.00 & \textbf{0.00} & 58.17 & 26.25 \\
\bf{FakeIDet2} & \textbf{4.53} & 1.45 & \textbf{7.17} & \textbf{3.99} \\
\bottomrule
\end{tabular}
\caption{Performance results for the fine-tuned DINOv2 (Entire \gls{id}), FakeIDet \cite{munoz2025exploring}, and proposed FakeIDet2 in terms of \gls{eer}. Results are achieved using the evaluation set of FakeIDet2-db. The Non-Anonymized configuration (Entire \gls{id}) simulates current scenarios in literature.}
\label{tab:patch_comp_performance}
\end{table}

In the 128$\times$128 configuration, we observe that FakeIDet2 performs better than FakeIDet in detecting fake IDs of type screen (1.23\% vs. 7.10\% \gls{eer}) and composite (2.09\% vs. 54.08\% \gls{eer}), and remains close to FakeIDet in detecting print fake IDs (0.12\% vs. 0.00\% \gls{eer}). Furthermore, we observe that for the physical composite fake IDs, the performance of FakeIDet drops to 54.08\% \gls{eer}, while FakeIDet2 accurately detects these types of fake IDs using patch feature extraction and fusion, achieving a 2.09\% \gls{eer}. The performance drop is attributed to the way FakeIDet computes the document level score. Given all patches from a real or fake \gls{id}, it calculates a score per individual patch and fuses them all together using a simple average, which yields a final score at the document level. FakeIDet is reliable if the \gls{pais} considered have the same patterns in the whole image. For example, in the case of a screen or print fake \gls{id}, it has the same properties throughout the whole \gls{id} (that is, traces or fingerprints of the fake \gls{id}), but this is not true for physical composite fake IDs, as the \gls{id} is mostly real with the exception of altered sections, failing to detect these types of fake ID properly. Finally, in the 64$\times$64 configuration, we can observe a similar trend, although performance seems to drop a bit in both FakeIDet and FakeIDet2.

These results prove the technical contributions of FakeIDet2 over FakeIDet \cite{munoz2025exploring}, i.e., Patch Embedding Extractor and
the Patch Embedding Fusion, that not only allows detecting attacks where the entire \gls{id} is a fake (i.e., print or screen), but also attacks where small sections are fake (i.e., composite). Given that the performance using the 64$\times$64 patch size configuration is close to the 128$\times$128 patch size, we selected this patch size for further experiments, as it increases privacy in real-world settings, which is the main purpose of the study.

\subsection{Anonymization vs. Performance}
\label{sub:anon_vs_performance}

In this section, we analyze the impact on performance when removing sensitive information from an \gls{id}. To do so, we train FakeIDet2 in the 64$\times$64 patch size configuration at all anonymization levels, and evaluate the performance using Non-Anonymized data. This scenario simulates realistic real-world settings in which ID Holders have some restrictions when sharing IDs to AI Researchers in order to train the fake ID detectors. Table \ref{tab:anon_degree_performance} shows the performance of FakeIDet2 in terms of \gls{eer} (\%) for the different anonymization levels and attacks. A particular trend can be observed: the fake ID detection errors (per attack and overall) increase as we reduce the amount of sensitive information, which shows that FakeIDet2 performance depends on the amount of sensitive information available in the training stage (3.99\% in Non-Anonymized vs. 8.64\% in Pseudo-Anonymized vs. 17.94\% in Fully-Anonymized). Even with this drop in terms of performance, we observe that the Pseudo-Anonymized 64$\times$64 configuration performs fairly well despite having seen only partial information of the \gls{id}, which supports our proposed hypothesis that training a model not using all the sensitive information of the \gls{id} may generate a small loss in performance as a trade-off. Given that this preliminary analysis confirmed that training with Pseudo-Anonymized data using 64$\times$64 patch size remains competitive, we decided to use this configuration for the rest of our experiments, as it is balanced in terms of privacy and performance, which is the purpose of the study. 

\begin{table}[t]
\centering
\begin{tabular}{lcccc}
\toprule
\textbf{Model} & \textbf{\makecell{Screen}} & \textbf{\makecell{Print}} & \textbf{\makecell{Composite}} & \textbf{All} \\
\midrule
Non-Anonymized & \textbf{4.53} & \textbf{1.45} & \textbf{7.17} & \textbf{3.99} \\
Pseudo-Anonymized & 5.01 & 4.89 & 16.29 & 8.64 \\
Fully-Anonymized & 8.98 & 5.77 & 26.54 & 17.94\\
\bottomrule
\end{tabular}
\caption{Performance of FakeIDet2 in terms of \gls{eer} (\%) when training over the different anonymization levels for the 64$\times$64 patch size configuration. Results are achieved using the final evaluation of FakeIDet2-db, considering Non-Anonymized \gls{ids}.}
\label{tab:anon_degree_performance}
\end{table}

\subsection{Leave-One-Attack-Out Scenario}
\label{sub:leave_one_attack_out}

\begin{table}
    \centering
    \begin{tabular}{lcccc}
        \toprule
        \textbf{Training} & \textbf{Screen} & \textbf{Print} & \textbf{Composite} & \textbf{All} \\
        \midrule
        No-Screen & 36.00 & 1.00 & 7.17 & 19.93 \\
        No-Print & 3.05 & 6.77 & 12.25 & 7.97 \\
        No-Composite & 2.02 & 0.00 & 50.96 & 28.24 \\
        \midrule
        All-Attacks & 5.01 & 4.89 & 16.29 & 8.64 \\
        \bottomrule
    \end{tabular}
    \caption{Performance of FakeIDet2 in terms of \gls{eer} (\%) for the leave-one-attack-out scenario, i.e., leaving one of the \gls{pais} out from the training dataset. These results are obtained evaluating with all \gls{pais}, including the one left out in each case.}
    \label{tab:leave_one_attack_out}
\end{table}

Table \ref{tab:leave_one_attack_out} provides the results of FakeIDet2 when different attacks are left out of the training data. For reference, we include \gls{eer} for FakeIDet2 when trained using all \gls{pais} available (All-Attacks). Based on our previous findings, we consider for training FakeIDet2 the Pseudo-Anonymized 64$\times$64 patch configuration, due to privacy reasons. The final evaluation is carried using Non-Anonymized 64$\times$64 patch configuration. 

The worst results in terms of \gls{eer} are obtained when composite attacks are left out, with a 28.24\% when evaluating all types of attacks. This shows that the physical composite attacks included in our new FakeIDet2-db are the harder attacks for our model to detect, with 50.96\% \gls{eer}. It is also interesting to note that our screen attacks seem difficult to detect when they are out of training, with 36.00\% \gls{eer}. Finally, in the case of the print attacks, FakeIDet2 performs much better compared to detecting the other attacks with just a 6.77\% in terms of \gls{eer} when only detecting print attacks, although a little caveat should be considered. As covered in Sec. \ref{sec:proposed_db}, our physical composite attacks are created by using small crops from print attacks and lying them over real \gls{ids}. We argue that FakeIDet2 may be leveraging those learned features from composite patches to predict print attacks appropriately, giving competitive results that even improve the performance when the whole dataset is available (8.64\% \gls{eer} vs 7.97\% \gls{eer}). 

Finally, an interesting phenomenon is observed when training FakeIDet2 leaving one attack out: the fake detector seems to ``specialize'' in detecting the attacks that are used in training, since the \gls{eer} becomes even lower compared to the case of training using all attacks. For example, when leaving the composite attacks, FakeIDet2 obtains an \gls{eer} of 2.02\% when detecting print attacks, compared to 5.01\% which is obtained when trained with all attacks. This can be expected, as when the number of classes is smaller, the data distribution to model by FakeIDet2 becomes simpler, hence becoming better at detecting the only attacks presented at training time.

\subsection{Leave-One-Sensor-Out Scenario}
\label{sub:leave_one_sensor_out}

\begin{table}
    \centering
    \begin{tabular}{lcccc}
        \toprule
        \textbf{Model} & \textbf{Screen} & \textbf{Print} & \textbf{Composite} & \textbf{All} \\
        \midrule
        No-iPhone & 22.12 & 30.98 & 30.63 & 22.61 \\
        No-Xiaomi & 3.96 & 1.94 & 14.24 & 8.30 \\
        No-Redmi & 8.07 & 1.94 & 7.13 & 8.97 \\ 
        \midrule
        All-Devices & 5.01 & 4.89 & 16.29 & 8.64 \\
        \bottomrule
    \end{tabular}
    \caption{Performance of FakeIDet2 in terms of \gls{eer} (\%) for the leave-one-sensor-out scenario, i.e., leaving one of the acquisition devices out from the training dataset. These results are obtained evaluating with data from all the acquisition devices in the Non-Anonymized, 64$\times$64 patch configuration including the one left out in each case.}
    \label{tab:leave_one_sensor_out}
\end{table}

Table \ref{tab:leave_one_sensor_out} provides the results of FakeIDet2 when different acquisition sensors are left out of the training data. For reference, we include \gls{eer} for FakeIDet2 when trained using all available devices (All-Devices). Based on our previous findings, we consider for training FakeIDet2 the Pseudo-Anonymized 64$\times$64 patch configuration, due to privacy reasons. The final evaluation is carried using Non-Anonymized 64$\times$64 patch configuration. 

As can be seen in the table, performance decreases when the highest quality sensor (i.e., iPhone 15) is not used in training but only in the final evaluation. This seems not to be happening in the rest of the scenarios, where images from the iPhone 15 are present in the training set. Given the different experiments that we run regarding this issue, our intuition on why this is happening relies on the fact that iPhone 15 cameras capture finer details that give more distinct features compared to Xiaomi's and Redmi's sensors. These fine-grain characteristics allow our FakeIDet2 to focus on traces that are more subtle, such as Moiré patterns on the screen attacks and the droplets of ink from the print and composite attacks, given that these attacks were printed with an Inkjet-kind printer. Furthermore, one of the main features of the AdaFace loss function is that the adaptive margins change depending on the image quality. When using images of high quality along with images of low or medium quality, the database is more diverse, and AdaFace helps the learning process by establishing larger margins for high quality images and lower for low quality images. If images of low quality are the majority of the database, the margins may not be as large as desired for proper separation between classes, and the features per each class may be overlapped in the embedded space, which are not helpful when forwarded into the Patch Embedding Fusion module, harming the performance.

\subsection{Cross-Database Scenario}
\label{sec:x_db_scenario}
This section evaluates the performance of FakeIDet2 trained through the selected configuration (i.e., Pseudo-Anonymized 64$\times$64 patch size) on different public databases in the literature. Concretely, we consider DLC-2021 \cite{polevoy2022document}, KID34K \cite{kid34k} and Benalcazar \textit{et al.} synthetic database from \cite{synth_id_card_db}. For this evaluation, we sample around 10,000 images from KID34K, balancing between \gls{pais} and \gls{id} owners, and approximately 30,000 images from several DLC-2021 videos, sampling in a similar way as in KID34K. DLC-2021 contains Spanish IDs, but they are from the first version, which are not seen during training FakeIDet2 as described in Sec. \ref{sub:train_settings}. Regarding the real IDs, as in those databases the ``real" IDs were created by the researchers under laboratory conditions, that is, they are not official IDs, we consider the aforementioned databases as attacks. Real IDs are extracted from the evaluation dataset of our FakeIDet2-db, similar to previous experiments, since they are official IDs. Our aim with this privacy-aware scenario is to assess the generalization ability of our proposed FakeIDet2 to unseen conditions, e.g., different \gls{pais}, templates, devices, etc. 

\begin{table}
    \centering
    \begin{tabular}{lccc}
    \toprule
    \textbf{PAI} & \textbf{DLC-2021} \cite{polevoy2022document} & \textbf{KID34K} \cite{kid34k} & \textbf{Benalcazar \textit{et al.}} \cite{synth_id_card_db} \\
    \midrule
    Screen & 5.02 & 13.41 & N/A\\
    HQ-Print & 10.28 & 18.26 & N/A\\
    Print & 12.45 & 4.99 & N/A\\
    Gray-Print & 10.60 & N/A & N/A\\
    Synthetic & N/A & N/A & 39.41 \\
    \midrule
    \makecell[c]{\textbf{All}} & \textbf{8.90} & \textbf{13.84} & \textbf{39.41} \\
    \bottomrule
    \end{tabular}
    \caption{Performance of FakeIDet2 in terms of \gls{eer} (\%) per type of attack and database. In order to facilitate the collaboration between ID Holders and AI Researchers, we consider the privacy-aware scenario where FakeIDet2 is trained using the Pseudo-Anonymized 64$\times$64 patch size configuration.}
    \label{tab:bf_is_bf}
\end{table}

Table \ref{tab:bf_is_bf} shows the performance of FakeIDet2 in terms of \gls{eer} (\%) per type of attack and database. In order to facilitate the collaboration between ID Holders and AI Researchers, we consider the privacy-aware scenario where FakeIDet2 is trained using the Pseudo-Anonymized 64$\times$64 patch size configuration. For DLC-2021 and KID34K we denote by ``HQ-Print'' the types of print attacks which are proposed as ``real" in each of the databases, and for Benalcazar \textit{et al.} we denote ``Synthetic'' the samples that have been created via generative methods trained from real samples. Furthermore, in case a database does not contain a specific attack, it is denoted as ``N/A''. 

As can be seen in Table \ref{tab:bf_is_bf}, the performance per \gls{pai} varies between databases. For example, FakeIDet2 is worse in screen attacks from the KID34K database compared to the DLC-2021, with 13.41\% vs. 5.02\% EER, respectively. This could be motivated by the amount of devices that were used to create screen attacks in KID34K, ranging from smartphones, tablets, laptops, and LCD monitors. A similar trend can be observed for the HQ-Print attacks, where the KID34K samples seem to confuse FakeIDet2 more (18.26\% vs 10.28\% EER). This may be induced by the variability in terms of acquisition devices (12 in KID34K vs. 2 in DLC-2021), with different specifications that create different traces or fingerprints in the images taken. Regarding print attacks, we see that our model performs better on the KID34K dataset than in DLC-2021 (4.99\% vs. 12.45\%). This might be produced because KID34K follows a similar approach to ours, i.e., print attacks are created after printed and then laminated. Finally, for the synthetic attacks included in Benalcazar \textit{et al.} database, FakeIDet2 is not able to generalize well to this unseen attack (39.41\% \gls{eer}). This result is expected due to the following reasons: \textit{i)} the data (both real and fake ID samples) used to train FakeIDet2 are only based on physical acquisitions using smartphone cameras, not synthetic data, and \textit{ii)} according to \cite{synth_id_card_db}, those synthetic samples were generated using \gls{gan} trained with official Chilean ID samples, which means that they share similar characteristics with real IDs that may confuse FakeIDet2. 

In conclusion, our proposed privacy-aware scenario in order to promote the collaboration between ID Holders and AI Researchers has provided very interesting results, despite the fact that our proposed FakeIDet2 is only trained through the Pseudo-Anonymized 64$\times$64 patch size configuration, using the development set of FakeIDet2-db. Under unseen physical attacks, FakeIDet2 is able to achieve 8.90\% \gls{eer} on DLC-2021 and 13.84\% on KID34K, as can be seen in Table \ref{tab:bf_is_bf}. We expect that these results can be enhanced in real-world settings, as the ID Holders own large-scale ID databases that can be exploited to improve the machine learning processes described here. 

\section{Conclusion}
\label{sec:conclusion}

In this article, we have explored a novel framework for privacy-aware fake \gls{id} detection, which bridges the existing gap between \glspl{id} Holders (e.g., governments, police, banks, etc.), which provide digital services to citizens and own large-scale datasets of real IDs, and AI Researchers, which have the experience and technology to develop fake \gls{id} detectors, but not the data. 

In order to advance in the field, we have first introduced FakeIDet2-db, a database that complies with the proposed framework, which contains more than 900K patches in different sizes (128$\times$128 and 64$\times$64), extracted from a total of 2,000 images of pseudo-anonymized and fully-anonymized \gls{ids}. Regarding real and fake \gls{ids}, to our knowledge, this is the first database that contains: \textit{i)} official real \gls{ids} captured under several conditions regarding camera sensor, illumination conditions and distance, and \textit{ii)} physical composite attacks. In addition, our database considers a large variability during the acquisition, e.g., using three different smartphones, and changes in illumination and height conditions. 

Together with FakeIDet2-db, we have presented FakeIDet2, a new method for the detection of fake \gls{id} that is sensitive to privacy and utilizes patches from \gls{ids} and varying levels of anonymization to protect sensitive information. We show that learning embeddings from patches extracted from real and fake \gls{ids} is a crucial step for effective fake \gls{id} detection. Considering the nature of the problem, especially for physical composite attacks, we have proposed a method that explicitly accounts for class imbalance using time-adaptive class weights \cite{fierrez18fusion} during learning and have compared it to other popular loss functions \cite{2022_PR_SetMargin_Morales}, achieving superior performance compared to traditional approaches (CosFace, ArcFace, and AdaFace). Using patch-level embeddings, we have incorporated a learnable fusion module based on multi-head self-attention \cite{attn_pool} to assign different weights to different patches within a single \gls{id}, producing an \gls{id}-level prediction that indicates whether the submitted \gls{id} is real or fake. Furthermore, we have evaluated the impact of different anonymization levels and patch sizes, showing that our FakeIDet2 remains competitive in privacy-aware scenarios, as compared to fully visible data settings (3.99\% in non-anonymized vs. 8.64\%
in pseudo-anonymized).

Finally, we have also introduced an extensive and reproducible public benchmark, which considers, in addition to our FakeIDet2-db, other public databases in the literature, containing unseen physical (KID34K \cite{kid34k}, DLC-2021 \cite{polevoy2022document}) and synthetic attacks (Benalcazar \textit{et al.} \cite{synth_id_card_db}) to assess performance in different data distributions that contain \gls{ids} from several demographic regions (i.e., Europe, Asia, and South America). 

The results achieved by FakeIDet2 (Pseudo-Anonymized 64$\times$64 patch size configuration) in the proposed benchmark under unseen conditions, 8.90\% \gls{eer} on DLC-2021 and 13.84\% on KID34K, promote our proposed privacy-aware scenario in order to facilitate the collaboration between ID Holders and AI Researchers in real-world settings. 

In future work, we will explore other foundation models \cite{oquab2023dinov2} and other learning architectures based on the patch-based principles shown in Fig.~\ref{fig:fakeidet-v2} and other non-patch and inpainting detection methods \cite{2023_IRL-Net_Ahmad} following ongoing benchmarks and competitions such as DeepID\footnote{\url{https://deepid-iccv.github.io/}}. Improving the accuracy of FakeIDet2 by exploiting the spatial context of patches (e.g., using spatial attention \cite{alcala25attzoom} and local quality measures \cite{2022_CSUR_FaceQsurvey_Torsten}) and a full layout analysis \cite{miguel25gnn} is also in our plans.

\section*{Acknowledgements}
\label{sec:acks}

Funding from INTER-ACTION (PID2021-126521OBI00 MICINN/FEDER), M2RAI (PID2024-160053OB-I00 MICIU/FEDER), Cátedra ENIA UAM-Veridas en IA Responsable (NextGenerationEU PRTR TSI-100927-2023-2), and PowerAI+ (SI4/PJI/2024-00062 Comunidad de Madrid and UAM).

% To print the credit authorship contribution details
%\printcredits

%% Loading bibliography style file
%\bibliographystyle{model1-num-names}
%\bibliographystyle{cas-model2-names}
\bibliographystyle{elsarticle-harv} 

% Loading bibliography database
\bibliography{cas-refs}

% Biography
%\bio{}
% Here goes the biography details.
%\endbio

%\bio{pic1}
% Here goes the biography details.
%\endbio

\end{document}